\documentclass[12pt]{article}
\usepackage{makeidx}
\usepackage{amssymb}
\usepackage{amsfonts}
\usepackage{amsmath}
\usepackage{graphicx}
\usepackage{setspace}
\usepackage{authblk}
\usepackage{units}
\usepackage{appendix}
\usepackage{xparse}
\usepackage{cite}
\usepackage[left=2.1cm,top=2.5cm,right=2.1cm]{geometry}

\begin{document}
	
	\title{Calculations of vacuum mean values of spinor field current and
		energy-momentum tensor in a constant electric background}
	
	\author{A.I. Breev$^{1}$\thanks{
			breev@mail.tsu.ru}, S.P. Gavrilov$^{1,2}$\thanks{
			gavrilovsergeyp@yahoo.com; gavrilovsp@herzen.spb.ru}, and D.M. Gitman$^{4,3,1}$
		\thanks{
			dmitrygitman@hotmail.com} \\
		{\normalsize $^{1}$ Department of Physics, Tomsk State University, Tomsk
			634050, Russia.}\\
		{\normalsize $^{2}$ Department of General and Experimental Physics, }\\
		{\normalsize Herzen State Pedagogical University of Russia,}\\
		{\normalsize Moyka embankment 48, 191186 St. Petersburg, Russia;}\\
		{\normalsize $^{3}$ P.N. Lebedev Physical Institute, 53 Leninsky prospekt,
			119991 Moscow, Russia;}\\
		{\normalsize $^{4}$ Institute of Physics, University of Sao Paulo, CEP
			05508-090, Sao Paulo, SP, Brazil; }
		}
	
	\maketitle

\begin{abstract}
In the framework of strong-field QED with $x$-steps, we study vacuum mean
values of the current density and energy--momentum tensor of the quantized
spinor field placed in the so-called $L$-constant electric
background. The latter background can be, for example, understood
as the electric field confined between capacitor plates, which are separated
by a sufficiently large distance $L$. First, we reveal peculiarities of
nonperturbative calculating of mean values in strong-field QED with $x$
-steps in general and, in the $L$-constant electric field, in particular. We
propose a new renormalization and volume regularization procedures that are
adequate for these calculations. We find necessary representations for
singular spinor functions in the background under consideration.
With their help, we calculate the above mentioned vacuum means. In the
obtained expressions, we show how to separate global contributions due to
the particle creation and local ones due to the vacuum polarization. We
demonstrate how these contributions can be related to the renormalized
effective Heisenberg--Euler Lagrangian.
\end{abstract}

\section{Introduction}

In QED with strong electric-like external fields (strong-field QED in what
follows) there exists the so-called vacuum instability due to the effect of
real particle creation from the vacuum caused by the external fields (the
so-called Schwinger effect \cite{Schwi51}). A number of publications,
reviews and books are devoted to this effect itself and to developing
different calculation methods in theories with unstable vacuum, see Refs. 
\cite{Nikis79,General,Greiner,FGS91,ruffini,GelTan16,Adv-QED22} for a
review. In strong-field QED, nonperturbative (with respect to
strong external fields) methods are well-developed for two classes of
external backgrounds, namely for the so-called $t$-electric potential steps 
($t$-steps) and $x$-electric potential steps ($x$-steps). $t$-steps represent
uniform time-dependent external electric fields that are switched on and off
at the initial and the final time instants, respectively whereas $x$-steps
represent time-independent external electric fields of constant direction
that are concentrated in restricted space areas. The latter fields can also
create particles from the vacuum, the Klein paradox is closely related to
this process \cite{Klein27,Sauter31a,Sauter-pot}. A general nonperturbative
formulation of strong-field QED with $t$-steps was developed many years ago
in Refs. \cite{Gitma77}. The study of particle creation due to the $x$-steps
began early in the framework of relativistic quantum mechanics, see Ref. 
\cite{DomCal99} for a review. However, until recently a consistent
quantum field theory (QFT) has not been completed. Only a short
time ago a nonperturbative formulation of strong-field QED with $x$-steps
was developed in Refs. \cite{GavGi16,GavGi20}. In the framework of
strong-field QED with $x$-steps calculations of particle creation effect
were presented in Refs. \cite
{L-field,mag-cr13,GavGitSh16,GavGitSh17,GavGitSh19,AdGG20,mag-cr21}. In both
relativistic quantum mechanics and strong-field QED the possibility of
nonperturbative calculations is based on the existence of specific exact
solutions (\textrm{in}- and \textrm{out}-solutions) of the Dirac equation.
In strong-field QED (in explicit in the relativistic quantum mechanics as
well), it is assumed that quantum processes under consideration do not
affect significantly classical external fields, the back-reaction is
supposed to be small. Nevertheless, it is well-understood that, in
principle, the back-reaction must be calculated, at any rate, to estimate
limits of the applicability of obtained results. It is also clear that the
back-reaction may be strong namely for external backgrounds that can violate
the vacuum stability. Here we have to say that studying the vacuum
instability, one usually calculates the number density of particles created
from the vacuum. In some cases, this allows one to make phenomenological
conclusions about the back-reaction; see, e.g., \cite{KluMotEis98}. However,
a complete study of the back-reaction is related to calculating mean values
of the current density and the energy-momentum tensor (EMT) of the charge
matter field. In strong-field QED with $t$-steps such a study was performed
in Refs. \cite{Gav06,GavGi08,GavGitY12}. In particular, it was demonstrated
that the effect of particle-creation is precisely the main reason for the
change of the energy of the matter. Making a comparison between the change
of the energy density of the charged matter and the energy density of the
external electric field, which is responsible for this change, restrictions
on the intensity of an external field and its duration were found, see Ref. 
\cite{GavGi08b}.

In the present article, in the framework of strong-field QED with $x$-steps,
we study vacuum mean values of the current density and EMT of the Dirac
field in a constant external electric field confined between capacitor
plates, which are separated by a sufficiently large
distance $L$. In earlier publications such a field is conditionally called 
$L$-constant electric field. In the limiting case $L\rightarrow \infty$ this 
field can be considered as a regularization of the constant
uniform electric field. In the obtained results, we demonstrate how to
separate contributions due to the global effects of particle production from
the local effect due to the vacuum polarization.Some relations with the
Heisenberg-Euler Lagrangian are established.

The paper is organized as follows. In Section \ref{S2a}, we describe
peculiarities of calculating mean values in strong-field QED with $x$-steps
and, in particular, in the $L$-constant electric field, in Section \ref{S2}.
In Section \ref{S4}, we refine the volume regularization procedure with
respect to the time-independent inner product on the $t$-constant
hyperplane and find necessary representations of singular spinor functions
in the electric field under consideration. In Section \ref{S5}, we calculate
directly the vacuum mean values of current density and EMT. In the obtained
expressions, we separate contributions due to the particle creation and due
to the vacuum polarization. We demonstrate how the latter contributions can
be derived by the help of the Heisenberg--Euler Lagrangian. In the last
section \ref{SConcl}, we summarize and discuss the main results. Some useful
technical details are placed in the Appendices.

In our consideration, we use the relativistic units $\hslash =c=1$ in which
the fine structure constant is $\alpha =e^{2}/c\hslash =e^{2}$.

\section{Mean values in strong-field QED with $x$-steps\label{S2a}}

We consider quantum and classical fields in $d=D+1$ dimensional Minkowski
space-time and use coordinates $X$, 
\begin{equation*}
	X=\left( X^{\mu },\ \mu =0,1,\ldots ,D\right) =\left( t,\mathbf{r}\right) ,\
	\ X^{0}=t,\ \ \mathbf{r}=\left( X^{1},\ldots ,X^{D}\right) ,\ \ x=X^{1}\ .
\end{equation*}
for their parametrization. We assume that the basic Dirac particle is an
electron with the mass $m$ and the charge $-e$, $e>0$, and the positron is
its antiparticle. The $x$-step ($L$-constant electric field) is given by
zero component of electromagnetic potential $A_{0}\left( x\right)$ that
depends on the coordinate $x$. The corresponding electric field $E\left(
x\right) =-\partial _{x}A_{0}\left(x\right) >0$ is directed
along the $x$ axis in the positive direction and is confined in the region $
S_{\mathrm{int}}=\left( x_{\mathrm{L}},x_{\mathrm{R}}\right) $, where $x_{
\mathrm{L}}<0$ and $x_{\mathrm{R}}>0$. The potential energy of an electron
is $U\left( x\right) =-eA_{0}\left( x\right) $, and $\partial _{x}U\left(
x\right) >0$ if $x\in S_{\mathrm{int}}$, and is constant outside the region $
S_{\mathrm{int}}$, $U\left( x\right) =U_{\mathrm{L}}$ if $x<x_{\mathrm{L}}$
and $U\left( x\right) =U_{\mathrm{R}}$ if $x>x_{\mathrm{R}}$. The field
accelerates the electrons along the axis $x$ in the negative direction and
the positrons along the axis $x$ in the positive direction. The $x$-step can
create particles from the vacuum if the magnitude of its potential energy is
sufficiently large, $\Delta U>2m$. Such a $x$-step is called critical. In
Refs. \cite{GavGi16,GavGi20} it was developed an approach that allows one to
calculate nonperturbatively effects of the vacuum instability in the
presence of $x$-steps (the above mentioned in the Introduction strong-field
QED with $x$-steps). It is clear that the process of pair creation is
transient. Nevertheless, the condition of the smallness of
backreaction shows there is a window in the parameter range of $E$ and a
time duration of its existence where the constant field approximation is
consistent \cite{GavGi08b}. Physically, one can believe that the
electric field of an $x$-step may be considered as a part of a
time-dependent inhomogeneous electric field $\mathbf{E}_{\mathrm{
pristine}}\left(X\right) $ directed along the $x$-direction,
{\large \ }which was switched on very fast before a time instant $t_{\mathrm{
in}}$, by this time it had time to spread to the whole region $S_{\mathrm{int
}}$. Then it was switched off very fast just after a time instant$
t_{\mathrm{out}}=t_{\mathrm{in}}+T$. We stress that the field $
\mathbf{E}_{\mathrm{pristine}}\left(X\right) $ is equal to $E\left(
x\right)$ from $t_{\mathrm{in}}$ to $t_{\mathrm{out}}$, 
considered in the region $S_{\mathrm{int}}$, acts as a constant field $E$
during the sufficiently large (macroscopic) period of time $T$, 
\begin{equation}
	T\gg \left( eE\right) ^{-1/2}\max \left\{ 1,m^{2}/eE\right\} \ .  
	\label{m27}
\end{equation}

We note, that there exist time-independent observables in the presence of
critical $x$-steps. The pair-production rate and the flux of created
particles are constant during the time\ $T$ and main contributions to the
latter quantities are independent from fast switching-on and -off effects if
Eq. (\ref{m27}) holds true. This statement is confirmed by results
obtained in considering exactly solvable cases with $t$-steps 
\cite{GavG96a,AdoGavGit17,AdoGavGit18,AdoFerGavGit18} and by numerical
calculations; see, e.g. \cite{KluMotEis98}\footnote{
Note that the pair-production rate per unit volume due to homogeneous fields
($x_{\mathrm{L}}\rightarrow -\infty $, $x_{\mathrm{R}}\rightarrow \infty$)
of given average intensity is equal to or higher than that for the case of
a finite width $x_{\mathrm{R}}-x_{\mathrm{L}}$; see Ref. \cite{L-field}.}
. Neglecting contributions of the fast switching-on and -off
effects, one can use in calculations instead of the true initial and final
vacua that existed before the time $t_{\mathrm{in}}$ and after the time $t_{
\mathrm{out}}$ some time-independent vacua $\left\vert 0,\mathrm{in}
\right\rangle$ and $\left\vert 0,\mathrm{out}\right\rangle$ respectively,
see Refs. \cite{GavGi16,GavGi20}.

In the case of the $L$-constant electric field we have $E\left( x\right) =E$
and $U\left( x\right) =eEx$ in the region $S_{\mathrm{int}}$ and we choose
that $x_{\mathrm{L}}=-L/2$ and $x_{\mathrm{R}}=L/2$. Its magnitude is 
\begin{equation*}
	\Delta U=U_{\mathrm{R}}-U_{\mathrm{L}}=eEL>0\ .
\end{equation*}
The $L$-constant field produces constant fluxes of created from the vacuum
final particles during the time interval $T$. These particles
created as electron--positron pairs and leave field area $S_{\mathrm{int}}$,
wherein electrons are emitted to the region $S_{\mathrm{L}}$ on the left of 
$S_{\mathrm{int}}$ and positrons to the region $S_{\mathrm{R}}$ on the right
of $S_{\mathrm{int}}$. In these regions the created particles have constant
velocities in opposite directions, moving away from the area $S_{\mathrm{int}
}$. They form constant longitudinal currents and energy fluxes in the
regions $S_{\mathrm{L}}$\ and $S_{\mathrm{R}}$, respectively. Since the time
interval $T$ is chosen be macroscopic, one may believe
that, measuring characteristics of particles in the regions $S_{\mathrm{L}}$
 and $S_{\mathrm{R}}$, we are able to evaluate the effect of pair
creation in the area $S_{\mathrm{int}}$ for the time interval $T$.
As it follows from exact results \cite{L-field} in the case of the $L$
-constant field with a sufficiently large length $L$, 
\begin{equation}
	L\gg \left( eE\right) ^{-1/2}\max \left\{ 1,m^{2}/eE\right\} \ ,
	\label{L-large}
\end{equation}
one can use semiclassical description. That is, $L$ is chosen be macroscopic
finite distance. In this description, created particles moving away from the
area $S_{\mathrm{int}}$ with sufficiently large absolute longitudinal
kinetic momenta. Outside the area $S_{\mathrm{int}}$, polarization effects
are absent, therefore, this is the final particle that will remain after the
field $E_{\mathrm{pristine}}\left( X\right)$ is turned off. They are
already formed as final particles in the field area. Thus, in the case of
the $L$-constant field with a sufficiently large length $L$, we are able to
measure characteristics of particles in the field area $S_{\mathrm{int}}$
on the plane $x=\mathrm{const.}$ for the time interval $T$.

We consider our theory in a large space-time box that has a spatial volume 
$V_{\bot}$ of the ($d-1$) dimensional hypersurface orthogonal to the
electric field direction and the time dimension $T$. From the latter point
of view, the vacuum mean values of the operators of physical quantities on
the plane $x=\mathrm{const}$ are defined as integrals over the area $V_{\bot
}$ of the plane $x=\mathrm{const}$ and the time interval $T$. Due to the
translational invariance of the external field in the $S_{\mathrm{int}}$,
all the mean values are proportional to the spatial volume $V_{\bot}$ and
the time interval $T$. In what follows, we consider mean values of the
operators $J^{\mu}(x)$ and $T_{\mu\nu}(x)$ with respect to both initial
and final vacua,
\begin{align}
& \left\langle J^{\mu }(x)\right\rangle _{\mathrm{in/out}}=-ie\text{\textrm{
tr}}\left[ \gamma ^{\mu }S_{\text{\textrm{in/out}}}^{c}(X,X^{\prime })\right]
|_{X=X^{\prime }}\ ,  \notag \\
& \left\langle J^{\mu }(x)\right\rangle ^{c}=-ie\text{\textrm{tr}}\left[
\gamma ^{\mu }S^{c}(X,X^{\prime })\right] |_{X=X^{\prime }}\ ;  \notag \\
& \left\langle T_{\mu \nu }(x)\right\rangle _{\mathrm{in/out}}=i\text{
\textrm{tr}}\left[ A_{\mu \nu }S_{\text{\textrm{in/out}}}^{c}(X,X^{\prime })
\right] |_{X=X^{\prime }}\ ,  \notag \\
& \left\langle T_{\mu \nu }(x)\right\rangle ^{c}=i\text{\textrm{tr}}\left[
A_{\mu \nu }S^{c}(X,X^{\prime })\right] |_{X=X^{\prime }}\ ,  \notag \\
& A_{\mu \nu }=\frac{1}{4}\left[ \gamma _{\mu }(P_{\nu }+P_{\nu }^{\prime
\ast })+\gamma _{\nu }\left( P_{\mu }+P_{\mu }^{\prime \ast }\right) \right]
\ ,  \label{m5.3b}
\end{align}
where $\gamma^{\mu}$ are $\gamma$-matrices in $d$ dimensions, 
\begin{equation*}
	\left[ \gamma ^{\mu },\gamma ^{\nu }\right] _{+}=2\eta ^{\mu \nu },\ \ 
	\eta_{\mu \nu }=\mathrm{diag}(1,-1,\ldots ,-1)\ .
\end{equation*}
In Eq. (\ref{m5.3b}) there appear the generalized causal \textrm{in--out}
propagator $S^{c}(X,X^{\prime})$, the so-called \textrm{in--in} propagator
 $S_{\text{\textrm{in}}}(X,X^{\prime })$, and \textrm{
out--out} propagator $S_{\text{\textrm{out}}}(X,X^{\prime })$ are used, 
\begin{align}
	& S^{c}(X,X^{\prime })=i\left\langle 0,\mathrm{out}\right\vert \hat{T}\hat{%
\Psi}(X)\hat{\Psi}^{\dag }(X^{\prime })\gamma ^{0}\left\vert 0,\mathrm{in}%
\right\rangle c_{v}^{-1},\;c_{v}=\langle 0,\mathrm{out}|0,\mathrm{in}\rangle
\ ,  \notag \\
	& S_{\text{\textrm{in}}}^{c}(X,X^{\prime })=i\left\langle 0,\mathrm{in}%
\right\vert \hat{T}\hat{\Psi}(X)\hat{\Psi}^{\dag }(X^{\prime })\gamma
^{0}\left\vert 0,\mathrm{in}\right\rangle \ ,  \notag \\
	& S_{\text{\textrm{out}}}^{c}(X,X^{\prime })=i\left\langle 0,\mathrm{out}%
\right\vert \hat{T}\hat{\Psi}(X)\hat{\Psi}^{\dag }(X^{\prime })\gamma
^{0}\left\vert 0,\mathrm{out}\right\rangle .  
	\label{m5.1}
\end{align}
Here $\hat{T}$ denotes the chronological ordering operation, $P_{\mu
}=i\partial _{\mu }+eA_{\mu }(X)$, $P_{\mu }^{\ast }=-i\partial _{\mu
}+eA_{\mu }(X)$, $\mathrm{tr}$ is denote the trace in the space, where 
$\gamma$-matrices are acting, and the Dirac Heisenberg operator 
$\hat{\Psi}(X)$ corresponds to the classical Dirac field $\psi(X)$. Here 
$\psi(X)$ is a $2^{[d/2]}$-component spinor (the brackets
stand for integer part of). The Dirac Heisenberg operator satisfies the
equal time canonical anticommutation relations
\begin{equation*}
	\left. \left[ \hat{\Psi}\left( X\right) ,\hat{\Psi}\left( X^{\prime }\right) 
	\right] _{+}\right\vert _{t=t^{\prime }}=0,\ \ \left. \left[ \hat{\Psi}
	\left( X\right) ,\hat{\Psi}^{\dagger }\left( X^{\prime }\right) \right]
	_{+}\right\vert _{t=t^{\prime }}=\delta \left( \mathbf{r-r}^{\prime }\right)
	\ .
\end{equation*}

It is clear that the vacuum polarization is a local effect, while the
concept of a particle has a clear meaning only after the electric field is
turned off, which in the case under consideration refers to those particles
that have left the field region. Nevertheless, it is natural to assume that
the created particles observed inside the field region near its boundaries 
$x_{\mathrm{L}}$ and $x_{\mathrm{R}}$ practically do not differ from those
observed outside this region and, therefore, represent the final particles.

In this article, our main task is to establish the relationship between the
matrix elements (\ref{m5.3b}) and the observable quantities that describe
the effects of vacuum polarization and particle production.

However, a number of important technical and principal questions still need
to be answered. The point is, that in the setting of problem considered in
Refs. \cite{GavGi16,GavGi20} did not consider local effects, produced by
electric field. In these works, it was assumed that the measurement of
particle fluxes through some surfaces $x=x_{\mathrm{meas}}^{\mathrm{L/R}}$, 
 $x_{\mathrm{meas}}^{\mathrm{L/R}}\in S_{\mathrm{L/R}}$ occurs at a
considerable distance from the field region $S_{\mathrm{int}}$ both in the
region $S_{\mathrm{L}}$ and in the region $S_{\mathrm{R}}$ during the
macroscopical time interval $T$.. This distance is assumed to be $cT$, which
is much larger than the extent of field, $x_{\mathrm{R}}-x_{\mathrm{L}}$. In
this case, during the time $T$ through surfaces 
$x=x_{\mathrm{meas}}^{\mathrm{L/R}}$, only those particles pass, which at the 
time of the beginning of the measurement were not farther from them than at a 
distance $cT$. Then the observed fluxes consist mainly of only one type of 
particles, namely, electrons in the region $S_{\mathrm{L}}$ and positrons in the 
region $S_{\mathrm{R}}$. Such a setting the problem allows you to neglect local
characteristics of the field and calculate the vacuum-to-vacuum transition
amplitude, mean differential and total numbers of created particles, mean
current and EMT of created particles for the case of arbitrary $x$-step. In
the case under consideration unlike the approach \cite{GavGi16,GavGi20} the
measurement of characteristics of particles is carried out in the field area 
$S_{\mathrm{int}}$, where fluxes consisting of both electrons and positrons
pass through any surface $x=\mathrm{const}$. This is a new type of task in
the framework of strong-field QED with $x$-steps, for which it is necessary
to re-establish the relationship between the duration of motion of particles
and a duration of observation. For this purpose, it is necessary to use a
certain regularization and renormalization of the parameters used. That is
why below we turn to a clarification of the physical meaning of these
parameters.

\section{Dirac field in the $L$-constant electric background \label{S2}}

The solutions of the Dirac equation with critical $x$-step are known in the
form of the stationary plane waves with given real longitudinal momenta
$p^{\mathrm{L}}$ and $p^{\mathrm{R}}$ in the regions $S_{\mathrm{L}}$ and 
$S_{\mathrm{R}}$, respectively. In this section we briefly recall some general 
features of these solutions established in Ref. \cite{GavGi16} and present 
necessary details for the case of the $L$-constant electric field; see Ref. 
\cite{L-field} for more details. We consider Dirac field in $d$ dimensional 
Minkowski space-time with coordinates $X$. A complete set of stationary plane 
waves has the following form:
\begin{eqnarray}
	&&\psi _{n}\left( X\right) =\left( \gamma ^{\mu }P_{\mu }+m\right) \Phi
	_{n}\left( X\right) ,\ \ \Phi _{n}\left( X\right) =\varphi _{n}\left(
	t,x\right) \varphi _{\mathbf{p}_{\bot }}\left( \mathbf{r}_{\bot }\right)
	v_{\chi ,\sigma }\ ,  \notag \\
	&&\varphi _{n}\left( t,x\right) =\exp \left( -ip_{0}t\right) \varphi
	_{n}\left( x\right) ,\;\ n=(p_{0},\mathbf{p}_{\bot },\sigma )\ ,
	\label{gav22} \\
	&&\mathbf{r}_{\bot }=\left( X^{2},\ldots ,X^{D}\right) ,\ \mathbf{p}_{\bot
	}=\left( p^{2},\ldots ,p^{D}\right) \ ,  \notag
\end{eqnarray}
where $v_{\chi,\sigma}$ with $\chi =\pm 1$ and $\sigma =(\sigma
_{1},\sigma _{2},\dots ,\sigma _{\lbrack d/2]-1})$, $\sigma _{j}=\pm 1$, is
a set of constant orthonormalized spinors satisfying the following
conditions:
\begin{equation*}
	\gamma ^{0}\gamma ^{1}v_{\chi ,\sigma }=\chi v_{\chi ,\sigma },\quad v_{\chi
	,\sigma }^{\dagger }v_{\chi ^{\prime },\sigma ^{\prime }}=\delta _{\chi
	,\chi ^{\prime }}\delta _{\sigma ,\sigma ^{\prime }}\ ,
\end{equation*}
In fact, functions (\ref{gav22}) correspond to states with given momenta 
$\mathbf{p}_{\bot }$ in the perpendicular to the axis $x$ direction. The
quantum numbers $\chi $ and $\sigma _{j}$ describe a spin polarization and
provide a convenient parametrization of the solutions. Since in ($1+1$) and 
$(2+1)$ dimensions ($d=2$, $3$) there are no any spin degrees
of freedom, the quantum numbers $\sigma$ are absent. Note that in ($2+1$)
dimensions, there are two nonequivalent representations for the 
$\gamma$-matrices which correspond to different fermion species parametrized by 
$\chi =\pm 1$ respectively. In $d$ dimensions, for any given momenta, there
exist only $J_{(d)}=2^{[d/2]-1}$ different spin states. One can see that
solutions (\ref{gav22}), which differ only by values of $\chi$, are
linearly dependent. Without loss of generality, we set $\chi=1$ and
introduce the notation $v_{\sigma }=v_{1,\sigma }$. The scalar functions 
$\varphi_{n}(x)$ obey the second-order differential equation:
\begin{equation}
	\left\{ \hat{p}_{x}^{2}-iU^{\prime }\left( x\right) -\left[ p_{0}-U\left(
	x\right) \right] ^{2}+\mathbf{p}_{\bot }^{2}+m^{2}\right\} \varphi
	_{n}\left( x\right) =0,\ \ \hat{p}_{x}=-i\partial _{x}\ .  
	\label{e3}
\end{equation}
Now we return to solving Eq. (\ref{e3}) in the area $x\in S_{\mathrm{int}}$.
It can be rewritten as follows:
\begin{equation}
	\left[ \frac{d^{2}}{d\xi ^{2}}+\xi ^{2}+i-\lambda \right] \varphi _{n}\left(
	x\right) =0,\ \ \xi =\frac{eEx-p_{0}}{\sqrt{eE}},\ \ \lambda =\frac{\pi
	_{\bot }^{2}}{eE}\ ,\ \ \pi _{\bot }=\sqrt{\mathbf{p}_{\bot }^{2}+m^{2}}\ .
	\label{L4}
\end{equation}
Note that $\pi_{0}\left( x\right) =p_{0}-eEx$ is kinetic energy an
electron. The general solution of Eq. (\ref{L4}) is completely determined by
an appropriate pair of linearly independent Weber parabolic cylinder
functions (WPCFs), either $D_{\rho }[(1-i)\xi ]$ and $D_{-1-\rho }[(1+i)\xi
] $ or $D_{\rho }[-(1-i)\xi ]$ and $D_{-1-\rho }[-(1+i)\xi ]$, where $\rho
=-i\lambda /2-1$.

We assume that corresponding potential step is sufficiently large, $\Delta
U=eEL\gg 2m$ (i.e., it is critical). In this case the field $E$ and leading
contributions to vacuum mean values can be considered as macroscopic
physical quantities.

In the case of critical steps, and, in particular, for the step under
consideration, there exist five ranges of quantum numbers $n$, 
$\Omega _{k}$, $k=1,\dots ,5$. We are interested in the Klein zone, the range 
$\Omega_{3},$ is defined by the inequalities 
$U_{\mathrm{L}}+\sqrt{eE\lambda }\leq p_{0}\leq U_{\mathrm{R}}-\sqrt{eE\lambda}$. 
Particle production from the vacuum takes place only in this range. We note that 
in the limit $L\rightarrow \infty $ the width of the Klein zone tends to the 
infinity.

For states with quantum numbers belonging to the Klein zone the $L$-constant
electric field can be considered as a regularization of a constant uniform
electric field. That is reason why such a field with a sufficiently large
length $L$, satisfying both condition (\ref{L-large}) and 
\begin{equation}
	\left[ \sqrt{eE}L\left( \sqrt{eE}L-2\sqrt{\lambda }\right) \right] ^{1/2}\gg
1\ ,  
	\label{L-large2}
\end{equation}
is of special interest. In what follows, we suppose that these conditions
hold true. Besides we assume that the additional condition
\begin{equation*}
	\sqrt{\lambda }<K_{\bot },\ \sqrt{eE}L/2\gg K_{\bot }^{2}\gg \max \left\{
	1,m^{2}/eE\right\}
\end{equation*}
takes place. Thus, in fact, we are going to consider the subrange $D$,
\begin{eqnarray}
	&&D\supset \Omega _{3}:\sqrt{\lambda }<K_{\bot },\;\left\vert
	p_{0}\right\vert /\sqrt{eE}<\sqrt{eE}L/2-K\ ,  \notag \\
	&&\sqrt{eE}L/2\gg K\gg K_{\bot }^{2}\gg \max \left\{ 1,m^{2}/eE\right\} \ ,
	\label{L25}
\end{eqnarray}
where $K$ and $K_{\bot }$ are any given numbers satisfying the condition 
(\ref{L25}). Namely in this subrange the pair creation is essential.

Solutions of the Dirac equation with well-defined left and right asymptotics
we denote as $_{\zeta }\psi _{n}\left( X\right) $ and $^{\zeta }\psi
_{n}\left( X\right)$,
\begin{eqnarray}
	&&\hat{p}_{x}\ _{\zeta }\psi _{n}\left( X\right) =p^{\mathrm{L}}\ _{\;\zeta
	}\psi _{n}\left( X\right) ,\ \ x\rightarrow x_{\mathrm{L}},\ \ \zeta =%
	\mathrm{sgn}(p^{\mathrm{L}})\ ,  \notag \\
	&&\hat{p}_{x}\ ^{\zeta }\psi _{n}\left( X\right) =p^{\mathrm{R}}\ ^{\zeta
	}\psi _{n}\left( X\right) ,\ \ x\rightarrow x_{\mathrm{R}},\ \ \zeta =%
	\mathrm{sgn}(p^{\mathrm{R}})\ ;  \notag \\
	&&p^{\mathrm{L}}=\zeta \sqrt{\left[ \pi _{0}\left( \mathrm{L}\right) \right]
	^{2}-\pi _{\bot }^{2}},\ \ p^{\mathrm{R}}=\zeta \sqrt{\left[ \pi _{0}\left( 
	\mathrm{R}\right) \right] ^{2}-\pi _{\bot }^{2}},\ \zeta =\pm \ ,  \notag \\
	&&\pi _{0}\left( \mathrm{L}\right) =p_{0}-U_{\mathrm{L}},\ \ \pi _{0}\left( 
	\mathrm{R}\right) =p_{0}-U_{\mathrm{R}}\ ,  
	\label{m3a}
\end{eqnarray}
where $\left\vert \pi _{0}\left( \mathrm{L}\right) \right\vert $ and 
$\left\vert \pi _{0}\left( \mathrm{R}\right) \right\vert $ are asymptotic
kinetic energies of an electron in the regions $S_{\mathrm{L}}$ and 
$S_{\mathrm{R}}$, respectively.

The solutions$\ _{\zeta }\psi _{n}\left( X\right)$ and$\ ^{\zeta }\psi
_{n}\left( X\right)$ describe particles with given momenta $p^{\mathrm{L}}$
as $x\rightarrow x_{\mathrm{L}}$ and $p^{\mathrm{R}}$ as 
$x\rightarrow x_{\mathrm{R}}$, respectively. One can see that the solutions 
$\ _{\zeta }\psi_{n}\left( X\right)$ and $\ ^{\zeta }\psi _{n}\left( X\right)$ 
have the form (\ref{gav22}) with functions $\varphi _{n}\left( x\right) $ denoted
here as $\ _{\zeta }\varphi _{n}\left( x\right) $ or $\ ^{\zeta }\varphi
_{n}\left( x\right)$ respectively. The latter functions have the following
asymptotics:
\begin{eqnarray*}
&&_{\zeta }\varphi _{n}\left( x\right) =\,_{\zeta }C\exp \left[ ip^{\mathrm{L
}}x\right] ,\quad x\rightarrow x_{\mathrm{L\ }}, \\
&&^{\zeta }\varphi _{n}\left( x\right) =\,^{\zeta }C\exp \left[ ip^{\mathrm{R
}}x\right] ,\quad x\rightarrow x_{\mathrm{R\ }}.
\end{eqnarray*}
Here$\ _{\zeta }C$ and $\ ^{\zeta }C$ are normalization constants.

It is supposed that all the solutions $\psi(X)$ are periodic
under transitions from one large space-time box that has a spatial volume 
$V_{\bot }$ to another. Under these suppositions, the inner product 
\begin{equation}
	\left( \psi ,\psi ^{\prime }\right) _{x}=\int_{V_{\bot }T}\psi ^{\dag
	}\left( X\right) \gamma ^{0}\gamma ^{1}\psi ^{\prime }\left( X\right) dtd%
	\mathbf{r}_{\bot }  
	\label{c3a}
\end{equation}
does not depend on $x$. The solutions $\ _{\zeta }\psi _{n}\left( X\right)$ 
and $\ ^{\zeta }\psi _{n}\left( X\right)$ satisfy the following
orthonormality relations on the $x=\mathrm{const}$ hyperplane: 
\begin{equation}
	\left( \ _{\zeta }\psi _{n},\ _{\zeta ^{\prime }}\psi _{n^{\prime }}\right)
	_{x}=\zeta \delta _{\zeta ,\zeta ^{\prime }}\delta _{n,n^{\prime
	}},\;\;\left( \ ^{\zeta }\psi _{n},\ ^{\zeta ^{\prime }}\psi _{n^{\prime
	}}\right) _{x}=-\zeta \delta _{\zeta ,\zeta ^{\prime }}\delta _{n,n^{\prime
	}}\ .  
	\label{c3b}
\end{equation}

In what follows, we will need two sets of solutions of equation for
the case $x_{\mathrm{L}}\rightarrow -\infty $ and 
$x_{\mathrm{R}}\rightarrow \infty $:
\begin{eqnarray}
	&&_{\;+}\varphi _{n}\left( x\right) =Y\ _{+}C\,D_{-1-\rho }[-(1+\mathrm{i}
	)\xi ],\ \ _{\;-}\varphi _{n}\left( x\right) =Y\ _{-}C\,D_{\rho }[-(1-
	\mathrm{i})\xi ]\ ,  \notag \\
	&&^{\;+}\varphi _{n}\left( x\right) =Y\ ^{+}C\,D_{\rho }[(1-\mathrm{i})\xi
	],\ \ ^{\;-}\varphi _{n}\left( x\right) =Y\ ^{-}C\,D_{-1-\rho }[(1+\mathrm{i}
)\xi ]\ ,  \notag \\
	&&\ ^{-\zeta }C=\,_{\zeta }C=\left( eE\right) ^{-1/2}e^{\pi \lambda /8}\left[
	\frac{\lambda }{2}(1+\zeta )+1-\zeta \right] ^{-1/2},\;Y=\left( V_{\bot
	}T\right) ^{-1/2}\ .  
	\label{set}
\end{eqnarray}
In the $V_{\bot }\rightarrow \infty$ and $T\rightarrow \infty$ limits one
has to replace the symbol $\delta _{n,n^{\prime }}$ in the normalization
conditions (\ref{c3b}) by quantity $\delta _{\sigma ,\sigma ^{\prime
}}\delta \left( p_{0}-p_{0}^{\prime }\right) \delta \left( \mathbf{p}_{\bot
}-\mathbf{p}_{\bot }^{\prime }\right)$ and to set $Y=\left( 2\pi \right)
^{-\left( d-1\right) /2}$ in Eq. (\ref{set}).

In the Klein zone, $\mathrm{in}$- and $\mathrm{out}$- solutions are: 
\begin{eqnarray}
	&&\mathrm{in-solutions:\ }\ \ _{-}\psi _{n}(X),\ ^{-}\psi _{n}(X)\ ,  \notag
	\\
	&&\mathrm{out-solutions:\ }\ \ _{+}\psi _{n}(X),\ ^{+}\psi _{n}(X)\ .
	\label{c17}
\end{eqnarray}
The solutions $\ ^{\zeta }\psi _{n_{3}}\left( X\right)$ describe electrons,
whereas the solutions $\ _{\zeta }\psi _{n_{3}}\left( X\right)$ describe
positrons.

The mutual decompositions of the solutions $_{\zeta }\psi _{n}\left(
X\right) $ and $^{\zeta }\psi _{n}\left( X\right) $ have the form:
\begin{eqnarray}
	&&^{\;\zeta }\psi _{n}\left( X\right) =\,_{+}\psi _{n}(X)g\left(
	_{+}\left\vert ^{\zeta }\right. \right) -\,_{-}\psi _{n}(X)g\left(
	_{-}\left\vert ^{\zeta }\right. \right) \ ,  \notag \\
	&&_{\;\zeta }\psi _{n}\left( X\right) =\,^{\;-}\psi _{n}\left( X\right)
	g\left( ^{-}\left\vert _{\zeta }\right. \right) -\,^{\;+}\psi _{n}\left(
	X\right) g\left( ^{+}\left\vert _{\zeta }\right. \right) \ ,  
	\label{rel01}
\end{eqnarray}
where expansion coefficients $g$ are defined by the relations:
\begin{equation}
	\left( \ _{\zeta }\psi _{n},\ ^{\;\zeta ^{\prime }}\psi _{n^{\prime }}\left(
	X\right) \right) _{x}=g\left( _{\zeta }\left\vert ^{\zeta ^{\prime }}\right.
	\right) \delta _{n,n^{\prime }},\ \ g\left( ^{\zeta ^{\prime }}\left\vert
	_{\zeta }\right. \right) =g\left( _{\zeta }\left\vert ^{\zeta ^{\prime
	}}\right. \right) ^{\ast }\ .  
	\label{c12}
\end{equation}
The coefficients $g$ satisfy the following unitary relations:
\begin{eqnarray*}
	&&\left\vert g\left( _{-}\left\vert ^{+}\right. \right) \right\vert
	^{2}=\left\vert g\left( _{+}\left\vert ^{-}\right. \right) \right\vert
	^{2},\;\left\vert g\left( _{+}\left\vert ^{+}\right. \right) \right\vert
	^{2}=\left\vert g\left( _{-}\left\vert ^{-}\right. \right) \right\vert ^{2}\
	, \\
	&&\frac{g\left( _{+}\left\vert ^{-}\right. \right) }{g\left( _{-}\left\vert
	^{-}\right. \right) }=\frac{g\left( ^{+}\left\vert _{-}\right. \right) }{%
	g\left( ^{+}\left\vert _{+}\right. \right) },\ \left\vert g\left(
	_{+}\left\vert ^{-}\right. \right) \right\vert ^{2}-\left\vert g\left(
	_{+}\left\vert ^{+}\right. \right) \right\vert ^{2}=1\ .
\end{eqnarray*}
The differential mean numbers of electrons and positrons from
electron-positron pairs created from the vacuum are equal and present the
number of created pairs,%
\begin{equation*}
N_{n}^{\mathrm{cr}}=\left\vert g\left( _{-}\left\vert ^{+}\right. \right)
\right\vert ^{-2}\ .
\end{equation*}%
The total number of pairs created from the vacuum $N^{\mathrm{cr}}$ is the
sum over the range $\Omega _{3}$ of the differential mean numbers 
$N_{n}^{\mathrm{cr}}$. Since the numbers $N_{n}^{\mathrm{cr}}$ do not depend on 
the spin polarization parameters $\sigma _{s}$, the sum over the spin
projections produces only the factor $J_{(d)}=2^{\left[ \frac{d}{2}\right]
-1}$. The sum over the momenta and the energy can be easily transformed into
an integral in the following way:
\begin{equation}
	N^{\mathrm{cr}}=\sum_{\mathbf{p}_{\bot },p_{0}\in \Omega _{3}}\sum_{\sigma
	}N_{n}^{\mathrm{cr}}=\frac{V_{\bot }TJ_{(d)}}{(2\pi )^{d-1}}\int_{\Omega
	_{3}}dp_{0}d\mathbf{p}_{\bot }N_{n}^{\mathrm{cr}}\ .  
	\label{L24}
\end{equation}

In the case of the $L$-constant electric field with a sufficiently large
length $L$, satisfying Eqs. (\ref{L-large}) and (\ref{L-large2}), functions 
(\ref{set}) have asymptotic expansions for $\left\vert \xi \right\vert \gg
\max \left\{ 1,\lambda \right\}$ (see, e.g. Ref. \cite{BatE53}) over the
wide range of energies $p_{0}$ for any given $\lambda $ of the subrange $D$
given by Eq. (\ref{L25}). In this subrange the quantity $N_{n}^{\mathrm{cr}}$
is almost constant and coincides with the well-known result in a constant
uniform electric field \cite{Nikis70a,Nikis70b,Nikis79}, 
\begin{equation}
	N_{n}^{\mathrm{cr}}\rightarrow N_{n}^{\mathrm{uni}}=e^{-\pi \lambda }\ .
	\label{L22}
\end{equation}
One can see that the formation interval over the $x$ for the mean numbers 
$N_{n}^{\mathrm{uni}}$ is the order of the length scale,
\begin{equation}
	\Delta l_{0}=\left( eE\right) ^{-1/2}\max \left\{ 1,\lambda \right\} \ .
	\label{L29a}
\end{equation}
Note that
\begin{equation*}
	N_{n}^{\mathrm{cr}}\sim \left\vert p^{\mathrm{R}}\right\vert \rightarrow 0,\
	\ N_{n}^{\mathrm{cr}}\sim \left\vert p^{\mathrm{L}}\right\vert \rightarrow
	0,\ \ \forall \lambda \neq 0\ ,
\end{equation*}
if $n$ tends to the boundary with either the range $\Omega _{2}$ $\left(
\left\vert p^{\mathrm{R}}\right\vert \rightarrow 0\right) $ or the range 
$\Omega _{4}$ $\left( \left\vert p^{\mathrm{L}}\right\vert \rightarrow
0\right)$ where the vacuum is stable.

In integral (\ref{L24}) $N_{n}^{\mathrm{cr}}$ plays the role of a cutoff
factor, that is why the main contribution is formed on the finite subrange 
$D$. Finally, we obtain:
\begin{equation}
	N^{\mathrm{cr}}=V_{\bot }Tn^{\mathrm{cr}},\;\;n^{\mathrm{cr}}=r^{\mathrm{cr}}
	\left[ L+\frac{O(K)}{\sqrt{eE}}\right] ,\;\;r^{\mathrm{cr}}=\frac{
	J_{(d)}\left( eE\right) ^{d/2}}{(2\pi )^{d-1}}\exp \left\{ -\pi \frac{m^{2}}{
	eE}\right\} \ .  
	\label{L27}
\end{equation}
Here $n^{\mathrm{cr}}$ is the total number density of created from the
vacuum pairs per unit of time and per unit of surface orthogonal to the
electric field direction.

Note that $n^{\mathrm{cr}}$ given by Eq. (\ref{L27}) is a function of the
field length $L$. The density $r^{\mathrm{cr}}=n^{\mathrm{cr}}/L$ is known
in the theory of pair creation in the constant uniform electric field as the
pair-production rate (see the $d$ dimensional case in Ref.~\cite{GavG96a}).

\section{Means of currents and EMT\label{S4}}

\subsection{Regularization\label{S3}}

Calculating some of the matrix elements considered above, one meets
divergences that indicate a need of a certain regularization. Below, we
consider such regularization and renormalization procedures for calculating
local quantities in strong-field QED with $L$-constant electric field. In
main, these procedures where formulated in Ref. \cite{GavGi16}, however,
here they are completed by some important and the necessary refinements.

In the case of the $L$-constant electric field under consideration, where
the distance $L$ between capacitor plates is sufficiently large, the plane
waves $_{\;\zeta }\psi _{n}\left( X\right)$ and 
$^{\;\zeta }\psi_{n}\left( X\right)$ can be identified by using one-particle mean 
currents and the energy fluxes in the field region $S_{\mathrm{int}}$, see Ref. 
\cite{L-field}. Thus, we can calculate the matrix elements (\ref{m5.3b}) inside
of the range $S_{\mathrm{int}}$. However, the explicit form of the singular
functions (\ref{m5.1}) depends on parameters of the volume regularization.
Due to physical reasons, these parameters are significantly different from
those proposed in the case when very wide regions $S_{\mathrm{L}}$ and 
$S_{\mathrm{R}}$ were used to measure fluxes of particles, see \cite{GavGi20}.
That is why below we turn to a clarification of the physical meaning of these 
parameters.

Stationary plane waves of type (\ref{m3a}) are usually used in potential
scattering theory, where they represent one-particle states with
corresponding conserved longitudinal currents. Such one-particle
consideration is consistent in all the ranges $\Omega_{k}$, excepting the
Klein zone $\Omega_{3}$. The technique developed in Ref. \cite{GavGi16}
does not need any refining in these ranges. Let us consider the range 
$\Omega_{3}$ where the strong-field QED consideration is essential. We note
that for our purposes it is sufficient to consider the subrange 
$D\supset\Omega_{3}$, which gives the main contribution to the vacuum 
instability.

The plane waves of the type (\ref{m3a}) are orthonormalized with respect to
the inner product (\ref{c3a}). To determine the time-independent
initial $\left\vert 0,\mathrm{in}\right\rangle$ and final $\left\vert 0, 
\mathrm{out}\right\rangle$ vacua and construct the corresponding 
\textrm{in}- and \textrm{out}-states in an adequate Fock space, we have to use a
time-independent inner product of solutions $\psi \left( X\right)$ and 
$\psi ^{\prime }\left( X\right)$ of the Dirac equation with the
field $E_{\mathrm{pristine}}\left( X\right)$ on a $t$ constant hyperplane. We 
recall that the periodic conditions are not imposed
in the $x$ direction. That is why, in contrast to the case of $t$-steps, the 
motion of particles in the $x$ direction is unlimited. Unlike the approach 
\cite{GavGi16,GavGi20} we assume that the large distance $L$ is not less then 
$cT$, where $T$ is an observation time $T$. In this case, one can ignore areas 
without the electric field and to believe that the part of the system under 
consideration causally related to the pair production process is situated inside 
the region $S_{\mathrm{int}}$. The corresponding particle states are represented 
by solutions given by Eqs. (\ref{gav22}) and (\ref{set}). For these reasons, we 
refine the volume regularization procedure used in Ref. \cite{GavGi16}, defining 
the time-independent inner product on the $t$-constant hyperplane as follows:
\begin{equation}
	\left( \psi ,\psi ^{\prime }\right) =\int_{V_{\bot }}d\mathbf{r}_{\bot
	}\int\limits_{-K^{\left( \mathrm{L}\right) }}^{K^{\left( \mathrm{R}\right)
	}}\psi ^{\dag }\left( X\right) \psi ^{\prime }\left( X\right) dx\ ,
	\label{t4}
\end{equation}
where the integral over the spatial volume $V_{\bot }$ is completed by the 
integral over the interval 
$\left[-K^{\left( \mathrm{L}\right) },K^{\left( \mathrm{R}\right) }\right]$ in 
the $x$ direction. Here $K^{\left( \mathrm{L/R}\right) }$ are some arbitrary
macroscopic but finite parameters of the volume regularization,
which are situated in the spatial area $S_{\mathrm{int}}$, 
$0<K^{\left( \mathrm{L}\right) }<\left\vert x_{\mathrm{L}}\right\vert$ and 
$0<K^{\left( \mathrm{R}\right) }<x_{\mathrm{R}}$. The length 
$K^{\left( \mathrm{R}\right) }+K^{\left( \mathrm{L}\right) }<L$ is sufficiently 
large,
\begin{equation*}
	K^{\left( \mathrm{R}\right) }+K^{\left( \mathrm{L}\right) }\gg 
	\Delta l_{0}\ ,
\end{equation*}
where $\Delta l_{0}$ is given by Eq. (\ref{L29a}).

Such an inner product is time-independent if solutions $\psi \left( X\right) 
$ and $\psi ^{\prime }\left( X\right)$ obey certain boundary conditions
that allow one to integrate by parts in Eq. (\ref{t4}) neglecting boundary
terms. The inner product (\ref{t4}) is conserved for such states. However,
considering solutions of the type (\ref{m3a}), which do not vanish at the
spatial infinity, we must accept some additional technical assumptions to
provide the time independence of the inner product (\ref{t4}). First of all,
we note that states with different quantum numbers $n$ are
independent, therefore decompositions of the vacuum matrix elements 
(\ref{m5.3b}) into the solutions with given $n$ do not contain interference
terms, see Appendix \ref{Ap1} for details. That is why it is
enough to consider Eq. (\ref{t4}) only for a particular case of solutions 
$_{\;\zeta }\psi _{n}\left( X\right) $ and $^{\;\zeta }\psi _{n}\left( X\right)$
with equal $n$. One can evaluate the principal value of integral (\ref{t4}) using 
relations (\ref{rel01}) and the asymptotic behavior of functions (\ref{set}) in 
the spatial regions where arguments of WPCF's are large, $\left\vert \xi
\right\vert \gg \max \left\{ 1,\lambda \right\}$, see Appendix \ref{Ap2}
for details. In this case the modulus of a longitudinal momentum is well
defined as, $\left\vert p_{x}\left( x\right) \right\vert =\sqrt{
\left[ \pi _{0}\left( x\right) \right] ^{2}-\pi _{\bot }^{2}}$. One
can see that the norms of the solutions $_{\;\zeta }\psi _{n}\left( X\right)$ and 
$^{\;\zeta }\psi _{n}\left( X\right) $ with respect to the inner product 
(\ref{t4}) are proportional to the macroscopically large parameters 
$\tau^{\left( \mathrm{L}\right) }$ and$\;\tau ^{\left( \mathrm{R}\right)}$,
\begin{equation*}
	\tau ^{\left( \mathrm{L}\right) }=K^{\left( \mathrm{L}\right) }/v^{\mathrm{L}
	},\;\tau ^{\left( \mathrm{R}\right) }=K^{\left( \mathrm{R}\right) }/v^{%
	\mathrm{R}}\ ,
\end{equation*}
where $v^{\mathrm{L}}=\left\vert p_{x}\left( x\right) /\pi
_{0}\left( x\right) \right\vert $ at $x=-K^{\left( \mathrm{L}\right) }$ and 
$v^{\mathrm{R}}=\left\vert p_{x}\left( x\right) /\pi _{0}\left( x\right)
\right\vert $ at $x=K^{\left( \mathrm{R}\right) }$ are absolute values of
the longitudinal velocities of particles. In the spatial regions of interest
where $\left\vert \xi \right\vert $ is large and the energy $\left\vert \pi
_{0}\left( x\right) \right\vert $ is much bigger then $\pi _{\bot }$, the
particles are moving almost parallel to the axis $x$, and the longitudinal
velocities $\left\vert p_{x}\left( x\right) /\pi _{0}\left( x\right)
\right\vert $ are ultrarelativistic at any $x$, such that $\left\vert
p_{x}\left( x\right) /\pi _{0}\left( x\right) \right\vert \rightarrow c$ 
($c=1$).

It was verified in Ref. \cite{GavGi16} that the following couples of
solutions are orthogonal with respect to the inner product (\ref{t4})
\begin{equation*}
	\left( _{\zeta }\psi _{n},^{\zeta }\psi _{n}\right) =0,\ \ 
	n\in \Omega_{3} 
\end{equation*}
if the parameters of the volume regularization 
$\tau ^{\left( \mathrm{L/R}\right) }$ satisfy the condition
\begin{equation}
	\tau ^{\left( \mathrm{L}\right) }-\tau ^{\left( \mathrm{R}\right) }=O
	\left(1\right) \ ,  
	\label{i8}
\end{equation}
where $O\left(1\right)$ are terms that are negligibly small in comparison
with the macroscopic quantities $\tau ^{\left( \mathrm{L/R}\right) }$. Thus, 
according to the physical interpretation given in latter reference,
the sets (\ref{c17}) represent \textrm{in}- and \textrm{out}-solutions,
which are linearly independent couples of complete on the 
$t$-constant hyperplane states with a given $n$. One can see that 
$\tau^{\left( \mathrm{L}\right)}$ and $\tau^{\left( \mathrm{R}\right)}$
are macroscopic times and they are equal,
\begin{equation*}
	\tau ^{\left( \mathrm{L}\right) }=\tau ^{\left( \mathrm{R}\right) }=\tau \ .
\end{equation*}

The $L$-constant field produces constant fluxes of created from the vacuum
final particles during the time interval $T$. These particles are
created with zero longitudinal kinetic momenta in a relatively small
formation interval $\Delta l_{0}$ given by Eq. (\ref{L29a}). After turning
into real particles electrons and positrons under the action of the electric
field move in opposite directions, the positrons move in the direction of
the electric field to the region $S_{\mathrm{R}}$, while the electrons in
the opposite direction to the region $S_{\mathrm{R}}$ and finally leave the
interval $\left[ -K^{\left( \mathrm{L}\right) },K^{\left( \mathrm{R}\right)}
\right] $. The time which is enough to these particles to reach one of the
hyperplane $x=-K^{\left( \mathrm{L}\right) }$ or $x=K^{\left( \mathrm{R}
\right) }$ varies from zero to the maximum possible time $2\tau $, which is
required by the ultrarelativistic particle to overcome the distance 
$K^{\left( \mathrm{R}\right) }+K^{\left( \mathrm{L}\right) }$. It is clear
that the kinetic energy (and the longitudinal kinetic momentum) of a
particle crossing these hyperplanes are proportional to the paths traveled
by the particles. Thus, the summation over the kinetic energies when
calculating fluxes of particles leaving the area between the hypersurfaces 
$x=-K^{\left( \mathrm{L}\right) }$ and $x=K^{\left( \mathrm{R}\right)}$ is
equivalent to the summation over the distances that these particles traveled
within the interval $\left[ -K^{\left( \mathrm{L}\right) },K^{\left( \mathrm{
R}\right) }\right]$ in the $x$-direction.

Under condition (\ref{i8}) the norms of the solutions on the $t$-constant
hyperplane are:
\begin{eqnarray}
	&&\left( \ _{\zeta }\psi _{n},\ _{\zeta }\psi _{n}\right) =\left( \ ^{\zeta
	}\psi _{n},\ ^{\zeta }\psi _{n}\right) =\mathcal{M}_{n}\ ,  \notag \\
	&&\mathcal{M}_{n}=2\frac{\tau }{T}\left\vert g\left( _{+}\left\vert
	^{-}\right. \right) \right\vert ^{2}\ \mathrm{if}\ \ n\in \Omega _{3}\ ,
	\label{i12}
\end{eqnarray}
where coefficients $g$ are defined by Eq. (\ref{c12}) and 
$\left\vert g\left( _{+}\left\vert ^{-}\right. \right) \right\vert ^{2}$ are 
given explicitly by Eq. (\ref{L22}). It is natural to assume that the observation
time $T$ (the time during which the observer registers flows of created
particles leaving the area between hyperplanes 
$x=-K^{\left( \mathrm{L}\right)}$ and $x=K^{\left( \mathrm{R}\right)}$) is 
equal to the maximal time $2\tau$,
\begin{equation}
	2\tau =T\ ,  
	\label{renorm}
\end{equation}
which is required for the created particles to leave the region with the
electric field. As we see in what follows, such a relation fixes the
proposed renormalization procedure. Thus, we find:
\begin{equation}
	\mathcal{M}_{n}=\left\vert g\left( _{+}\left\vert ^{-}\right. \right)
	\right\vert ^{2}\ \ \mathrm{if}\ \ n\in \Omega _{3}\ .  
	\label{i12b}
\end{equation}
In the case $L\rightarrow \infty $, one can consider the limit 
$V_{\bot}, K^{\left( \mathrm{L}/\mathrm{R}\right)}\rightarrow \infty$ to obtain
normalized solutions in the range $\Omega_{3}$ as follows:
\begin{equation*}
	\left( \ _{\zeta }\psi _{n},\ _{\zeta }\psi _{n^{\prime }}\right) =\left( \
	^{\zeta }\psi _{n},\ ^{\zeta }\psi _{n^{\prime }}\right) =\delta _{\sigma
	,\sigma ^{\prime }}\delta \left( p_{0}-p_{0}^{\prime }\right) \delta (%
	\mathbf{p}_{\bot }-\mathbf{p}_{\bot }^{\prime })\mathcal{M}_{n},\ \;\left(
	_{\zeta }\psi _{n},^{\zeta }\psi _{n^{\prime }}\right) =0\ ,
\end{equation*}
where the quantity $\mathcal{M}_{n}$ is given by Eq. (\ref{i12b}).

\subsection{Singular functions\label{S3b}}

We recall that in the general case, in theories with unstable vacuum, the
singular functions (\ref{m5.1}) do not coincide. The differences between the
functions $S_{\text{\textrm{in}}}^{c}(X,X^{\prime })$, 
$S_{\text{\textrm{out}}}^{c}(X,X^{\prime })$ and the causal propagator $S^{c}
(X,X^{\prime })$ are denoted by $S^{p}(X,X^{\prime })$ and $S^{\bar{p}}
(X,X^{\prime })$,
\begin{eqnarray}
	S^{p}(X,X^{\prime }) &=&S_{\mathrm{in}}^{c}(X,X^{\prime })-S^{c}(X,X^{\prime
	})\ ,  \notag \\
	S^{\bar{p}}(X,X^{\prime }) &=&S_{\mathrm{out}}^{c}(X,X^{\prime
	})-S^{c}(X,X^{\prime })\ .  
	\label{gav13a}
\end{eqnarray}

In the case of strong-field QED with $L$-constant electric field, all the
functions can be expressed as sums over the solutions, given by Eqs. 
(\ref{gav22}) and (\ref{set}), see Ref. \cite{GavGi16}. It can be seen that in
the case under consideration with $L\rightarrow \infty$, the main
contributions to the sums are due to the Klein zone. Taking this fact into
account, the singular functions can be represented as:
\begin{eqnarray}
	&&S^{c}(X,X^{\prime })=\theta (t-t^{\prime })\,S^{-}\left( X,X^{\prime
	}\right) -\theta (t^{\prime }-t)\,S^{+}\left( X,X^{\prime }\right) \ , 
	\notag \\
	&&S^{-}(X,X^{\prime })=i\sum_{n}\mathcal{M}_{n}^{-1}\ ^{+}\psi _{n}\left(
	X\right) g\left( ^{+}\left\vert _{-}\right. \right) g\left( ^{-}\left\vert
	_{-}\right. \right) ^{-1}\ ^{-}\bar{\psi}_{n}\left( X^{\prime }\right) \ , 
	\notag \\
	&&S^{+}(X,X^{\prime })=i\sum_{n}\mathcal{M}_{n}^{-1}\ _{-}\psi _{n}\left(
	X\right) g\left( _{-}\left\vert ^{+}\right. \right) g\left( _{+}\left\vert
	^{+}\right. \right) ^{-1}\ _{+}\bar{\psi}_{n}\left( X^{\prime }\right) \ ;
	\label{gav11b} \\
	&&S_{\text{\textrm{in/out}}}^{c}(X,X^{\prime })=\theta (t-t^{\prime })\,S_{%
	\mathrm{in/out}}^{-}\left( X,X^{\prime }\right) -\theta (t^{\prime }-t)\,S_{%
	\mathrm{in/out}}^{+}\left( X,X^{\prime }\right) \ ,  \notag \\
	&&S_{\mathrm{in/out}}^{-}(X,X^{\prime })=i\sum_{n}\mathcal{M}_{n}^{-1}\;\
	^{\mp }\psi _{n}\left( X\right) \ ^{\mp }\bar{\psi}_{n}\left( X^{\prime
	}\right) \ ,  \notag \\
	&&S_{\mathrm{in/out}}^{+}(X,X^{\prime })=i\sum_{n}\mathcal{M}_{n}^{-1}\;\
	_{\mp }\psi _{n}\left( X\right) \ _{\mp }\bar{\psi}_{n}\left( X^{\prime
	}\right) ,\ \bar{\psi}=\psi ^{\dagger }\gamma ^{0}\ ,  
	\label{gav11c}
\end{eqnarray}
where $\mathcal{M}_{n}$ is given by Eqs. (\ref{i12b}).

Using relations (\ref{rel01}), we represent the singular functions 
$S^{p}(X,X^{\prime })$ and $S^{\bar{p}}(X,X^{\prime })$
given by Eq. (\ref{gav13a}) as follows: 
\begin{eqnarray}
	&&S^{p}(X,X^{\prime })=i\sum_{n}\mathcal{M}_{n}^{-1}\ _{-}\psi _{n}\left(
	X\right) g\left( ^{-}\left\vert _{-}\right. \right) ^{-1}\ ^{-}\bar{\psi}
	_{n}\left( X^{\prime }\right) \ ,  \notag \\
	&&S^{\bar{p}}(X,X^{\prime })=-i\sum_{n}\mathcal{M}_{n}^{-1}\;^{+}\psi
	_{n}\left( X\right) g\left( _{+}\left\vert ^{+}\right. \right) ^{-1}\ _{+}
	\bar{\psi}_{n}\left( X^{\prime }\right) \ .  
	\label{gav13b}
\end{eqnarray}
We stress that both functions vanish in the absence of the vacuum
instability.

\section{Calculation of mean values in strong-field QED with L-constant
field \label{S5}}

\subsection{Pair-creation contributions\label{S51}}

With account taken of (\ref{gav13a}) the vacuum matrix elements, defined by
Eqs. (\ref{m5.3b}) and (\ref{m5.1}), can be represented as: 
\begin{eqnarray}
	&&\left\langle J^{\mu }(x)\right\rangle _{\mathrm{in}}=\mathrm{\mathrm{Re}}
	\left\langle J^{\mu }(x)\right\rangle ^{c}+\mathrm{\mathrm{Re}}\left\langle
	J^{\mu }(x)\right\rangle ^{p},\;\left\langle J^{\mu }(x)\right\rangle _{
	\mathrm{out}}=\mathrm{\mathrm{Re}}\left\langle J^{\mu }(x)\right\rangle ^{c}+
	\mathrm{\mathrm{Re}}\left\langle J^{\mu }(x)\right\rangle ^{\bar{p}}\ ,  \notag\\
	&&\left\langle J^{\mu }(x)\right\rangle ^{p,\bar{p}}=-ie\,\left. \mathrm{tr}
	\left[ \gamma ^{\mu }S^{p,\bar{p}}(X,X^{\prime })\right] \right\vert
	_{X=X^{\prime }};  \notag \\
	&&\left\langle T_{\mu \nu }(x)\right\rangle _{\mathrm{in}}=\left\langle 0,
	\mathrm{in}\right\vert T_{\mu \nu }\left\vert 0,\mathrm{in}\right\rangle
	=i\,\left. \mathrm{tr}\left[ A_{\mu \nu }S_{\mathrm{in}}^{c}(X,X^{\prime })
	\right] \right\vert _{X=X^{\prime }}\ =\mathrm{\mathrm{Re}}\left\langle 
	T_{\mu\nu }(x)\right\rangle ^{c}+\mathrm{\mathrm{Re}}\left\langle T_{\mu \nu
	}(x)\right\rangle ^{p}\ ,  \notag \\
	&&\left\langle T_{\mu \nu }(x)\right\rangle _{\mathrm{out}}=\left\langle 0,
	\mathrm{out}\right\vert T_{\mu \nu }\left\vert 0,\mathrm{out}\right\rangle
	=i\,\left. \mathrm{tr}\left[ A_{\mu \nu }S_{\mathrm{out}}^{c}(X,X^{\prime })
	\right] \right\vert _{X=X^{\prime }}=\left\langle T_{\mu \nu
	}(x)\right\rangle ^{c}+\left\langle T_{\mu \nu }(x)\right\rangle ^{\bar{p}}\
	,  \notag \\
	&&\left\langle T_{\mu \nu }(x)\right\rangle ^{p,\bar{p}}=i\,\left. \mathrm{tr
	}\left[ A_{\mu \nu }S^{p,\bar{p}}(X,X^{\prime })\right] \right\vert
	_{X=X^{\prime }}\ .  
	\label{4.1b}
\end{eqnarray}

One can see with help of Eqs. (\ref{gav11b}) and (\ref{gav13b}) that all the
quantities $\left\langle J^{\mu}(x)\right\rangle ^{c}$ and $\left\langle
T_{\mu \nu }(x)\right\rangle ^{c}$ are finite as $L\rightarrow \infty$,
whereas the current components 
$\left\langle J^{0}(x)\right\rangle ^{p,\bar{p}}$, 
$\left\langle J^{1}(x)\right\rangle ^{p,\bar{p}}$, 
$\left\langle T^{10}(x)\right\rangle ^{p/\bar{p}}$, and the diagonal components 
$\left\langle T_{\mu \mu }(x)\right\rangle ^{p,\bar{p}}$ of the EMT are
growing unlimited as $L$. That is why here we consider the components 
$\left\langle J^{\mu }(x)\right\rangle ^{p,\bar{p}}$ and 
$\left\langle T_{\mu\nu }(x)\right\rangle ^{p,\bar{p}}$ for the case of a large 
but finite $L$.

In representation (\ref{gav13b}) the factor $\mathcal{M}_{n}^{-1}=N_{n}^{
\mathrm{cr}}$ plays the role of a cutoff factor, that is why the main
contribution is formed on the finite subrange $D$ given by Eq.~(\ref{L25}).
That is why all the integrals over the momenta are finite. $N_{n}^{\mathrm{cr
}}$ is given by Eq. (\ref{L22}) in the subrange $D$ and does not depend on 
$p_{0}$. In the subrange $D$ and for large $L$, the integral over $p_{0}$ is
responsible for growing contributions as $L\rightarrow \infty$. That is why
the main contribution to the vacuum means under consideration are formed in
this subarea, such that it is enough to consider further the following
expressions 
\begin{eqnarray}
	&&\ \ S^{p}(X,X^{\prime })=i\sum_{n\in D}N_{n}^{\mathrm{cr}
	}g(^{-}|_{-})^{-1}\ _{-}\psi _{n}\left( X\right) \ ^{-}\bar{\psi}_{n}\left(
	X^{\prime }\right)  \notag \\
	&&\ =\frac{iV_{\bot }T}{(2\pi )^{d-1}}\int_{D}dp_{0}d\mathbf{p}_{\bot
	}\sum_{\sigma }N_{n}^{\mathrm{cr}}g(^{-}|_{-})^{-1}\ _{-}\psi _{n}\left(
	X\right) \ ^{-}\bar{\psi}_{n}\left( X^{\prime }\right) \ ,  \notag \\
	&&\ \ S^{\bar{p}}(X,X^{\prime })=-i\sum_{n\in D}N_{n}^{\mathrm{cr}}g\left(
	_{+}\left\vert ^{+}\right. \right) ^{-1}\;^{+}\psi _{n}\left( X\right) \ _{+}
	\bar{\psi}_{n}\left( X^{\prime }\right)  \notag \\
	&&\ =-\frac{iV_{\bot }T}{(2\pi )^{d-1}}\int_{D}dp_{0}d\mathbf{p}_{\bot
	}\sum_{\sigma }N_{n}^{\mathrm{cr}}g\left( _{+}\left\vert ^{+}\right. \right)
	^{-1}\ ^{+}\psi _{n}\left( X\right) \ _{+}\bar{\psi}_{n}\left( X^{\prime
	}\right) \   
	\label{m13.2b}
\end{eqnarray}
for the singular functions (\ref{gav13b}).

We are interested in the mean values under consideration inside the
capacitor, namely for $x\in S_{\mathrm{int}},$ where 
$\left\vert x\right\vert <L/2$. In the range $D$ for the given 
$x\sim x^{\prime}$ you can select subranges $D_{x}^{+}\subset D$ and 
$D_{x}^{-}\subset D$,
\begin{eqnarray}
	D_{x}^{+} &:&-\left[ eE\left( x-x_{\mathrm{L}}\right) -K\sqrt{eE}\right]
	<\pi _{0}\left( x\right) <-\sqrt{eE}K,\;\sqrt{\lambda }<K_{\bot }\;\mathrm{if 
	}\;\xi >0\ ,  \notag \\
	D_{x}^{-} &:&\sqrt{eE}K<\pi _{0}\left( x\right) <eE\left( x_{\mathrm{R}
	}-x\right) -K\sqrt{eE},\;\sqrt{\lambda }<K_{\bot }\;\mathrm{if}\;\xi <0\ ,
	\label{Dx}
\end{eqnarray}
where $\left\vert \pi _{0}\left( x\right) \right\vert$ is
sufficiently large. In these subranges, the functions 
$\ _{-}\psi _{n}\left(X\right)$ and 
$\ ^{-}\bar{\psi}_{n}\left(X^{\prime}\right)$ 
can be approximated by asymptotic forms of the WPCF's for big 
$\left\vert \xi\right\vert \sim \left\vert \xi ^{\prime }\right\vert >K$, 
where $\xi^{\prime }=\left. \xi \right\vert _{x\rightarrow x^{\prime }}$; e.g., 
see Ref. \cite{BatE53}. Note that $\left\vert \pi _{0}\left( x\right)
\right\vert $ is kinetic energy of a positron in $D_{x}^{+}$ and 
$\pi_{0}\left(x\right)$ is kinetic energy of an electron in $D_{x}^{-}$. Both
integration domains in Eq. (\ref{Dx}) are large enough, to provide the main
contribution to the integrals (\ref{m13.2b}).

Let us consider the case of 
$\pi _{0}\left( x\right) \sim \pi _{0}\left(
x^{\prime }\right) \in D_{x}^{+}$. By the help of Eq. (\ref{rel01}) we get:
\begin{eqnarray*}
	&&S^{p}(X,X^{\prime })=\nonumber\\
	&&-\frac{iV_{\bot }T}{(2\pi )^{d-1}}\int_{D}dp_{0}d
	\mathbf{p}_{\bot }\sum_{\sigma }N_{n}^{\mathrm{cr}}g(^{-}|_{-})^{-1}\left[ \
	^{+}\psi _{n}\left( X\right) g\left( ^{+}\left\vert _{-}\right. \right) -\
	^{-}\psi _{n}\left( X\right) g\left( ^{-}\left\vert _{-}\right. \right)
	\right] \ ^{-}\bar{\psi}_{n}\left( X^{\prime }\right) \ , \\
	&&S^{\bar{p}}(X,X^{\prime })=\nonumber\\
	&&-\frac{iV_{\bot }T}{(2\pi )^{d-1}}
	\int_{D}dp_{0}d\mathbf{p}_{\bot }\sum_{\sigma }N_{n}^{\mathrm{cr}}\ g\left(
	_{+}\left\vert ^{+}\right. \right) ^{-1}\ ^{+}\psi _{n}\left( X\right) \left[
	g\left( _{+}\left\vert ^{-}\right. \right) \ ^{-}\bar{\psi}_{n}\left(
	X^{\prime }\right) -\ ^{+}\bar{\psi}_{n}\left( X^{\prime }\right) g\left(
	_{+}\left\vert ^{+}\right. \right) \right] \ .
\end{eqnarray*}

In the case $X\sim X^{\prime }$, using asymptotics of WPCF's ,
given by Eq. (\ref{set}) and discarding negligibly small contributions from
the oscillating terms to the integral over $p_{0}$, we obtain:
\begin{eqnarray*}
	&&S^{p}(X,X^{\prime })\approx S_{+}^{p}(X,X^{\prime })=\nonumber\\
	&&\frac{iV_{\bot }T}{
	(2\pi )^{d-1}}\int_{D_{x}^{+}}dp_{0}d\mathbf{p}_{\bot }\sum_{\sigma }e^{-\pi
	\lambda }\ ^{-}\psi _{n}\left( X\right) \ ^{-}\bar{\psi}_{n}\left( X^{\prime
	}\right) =(\gamma P+m)\Delta _{+}^{p}(X,X^{\prime })\ , \\
	&&S^{\bar{p}}(X,X^{\prime })\approx S_{+}^{\bar{p}}(X,X^{\prime })=\nonumber\\
	&&\frac{
	iV_{\bot }T}{(2\pi )^{d-1}}\int_{D_{x}^{+}}dp_{0}d\mathbf{p}_{\bot
	}\sum_{\sigma }e^{-\pi \lambda }\ [^{+}\psi _{n}\left( X\right) \ ^{+}\bar{
	\psi}_{n}\left( X^{\prime }\right) ]=(\gamma P+m)\Delta _{+}^{\bar{p}
	}(X,X^{\prime })\ , \\
	&&\Delta _{+}^{p/\bar{p}}(X,X^{\prime })\sim\nonumber\\
	&&\frac{-i}{2\sqrt{eE}(2\pi
	)^{d-1}}\int_{D_{x}^{+}}dp_{0}d\mathbf{p}_{\bot }\xi ^{-1}\exp \left[ -\pi
	\lambda -ip_{0}(t-t^{\prime })+i\mathbf{p}_{\bot }(\mathbf{r}_{\bot }-
	\mathbf{r}_{\bot }^{\prime })\pm i\frac{\xi ^{2}-\xi ^{\prime 2}}{2}\right]
	\ .
\end{eqnarray*}
Now we consider the integration over the transversal momenta 
$\mathbf{p}_{\bot }$. In the subrange $D_{x}^{+}$, given by Eq. (\ref{Dx}), the 
domain of the variation of $\left\vert \mathbf{p}_{\bot }\right\vert$ is finite.
However, taking into account that the exponential $\exp (-\pi \lambda)$
plays the role of a cutoff factor, we can extend the limits of the domain to
infinity. As a result we have:
\begin{eqnarray}
	&&S_{+}^{p/\bar{p}}(X,X^{\prime })=(\gamma P+m)\Delta _{+}^{p/\bar{p}
	}(X,X^{\prime })\ ,  \notag \\
	&&\Delta _{+}^{p/\bar{p}}(X,X^{\prime })=-i\,h_{\perp }(\mathbf{r}_{\bot },
	\mathbf{r}_{\bot }^{\prime })\int_{x_{\mathrm{L}}+K/\sqrt{eE}}^{x-K/\sqrt{eE}
	}h_{\parallel }^{-/+}(x,\tilde{x})d\tilde{x},\ \ p_{0}=eE\,\tilde{x}\ , 
	\notag \\
	&&h_{\parallel }^{-/+}(x,\tilde{x})=\frac{1}{2\left( x-\tilde{x}\right) }
	\exp \left\{ -ip_{0}(t-t^{\prime })\mp \frac{i}{2}\left[ \xi (x)^{2}-\xi
	(x^{\prime })^{2}\right] \right\} \ .  \notag \\
	&&h_{\perp }(\mathbf{r}_{\bot },\mathbf{r}_{\bot }^{\prime })=\frac{\left(
	eE\right) ^{d/2-1}}{(2\pi )^{d-1}}\exp \left( -\frac{\pi m^{2}}{eE}-\frac{
	eE\,\left\vert \mathbf{r}_{\bot }-\mathbf{r}_{\bot }^{\prime }\right\vert
	^{2}}{4\pi }\right) \ .  
	\label{dip01}
\end{eqnarray}

Let us consider the case of $\pi _{0}\left(x\right) \sim \pi_{0}\left(
x^{\prime }\right) \in D_{x}^{-}$. In the same way as before, we can justify
that it is enough to consider further the following expressions for the
singular functions (\ref{gav13b}): 
\begin{eqnarray}
	&&S_{-}^{p/\bar{p}}(X,X^{\prime })=(\gamma P+m)\Delta _{-}^{p/\bar{p}
	}(X,X^{\prime }),\ \ \xi <-K\ ,  \notag \\
	&&\Delta _{-}^{p/\bar{p}}(X,X^{\prime })=+i\,h_{\perp }(\mathbf{r}_{\bot },
	\mathbf{r}_{\bot }^{\prime })\int_{x+K/\sqrt{eE}}^{x_{\mathrm{R}}-K/\sqrt{eE}
	}h_{\parallel }^{+/-}(x,\tilde{x})d\tilde{x},\ \ p_{0}=eE\,\tilde{x}\ .
	\label{dip02}
\end{eqnarray}

Now, we calculate the vacuum means values under consideration, which, for a
given $x$, are formulated by contributions from both domains $D_{x}^{+}$ 
and $D_{x}^{-}$,
\begin{eqnarray}
	\left\langle J^{\mu }(x)\right\rangle ^{p/\bar{p}} &=&\left\langle J^{\mu
	}(x)\right\rangle _{+}^{p/\bar{p}}+\left\langle J^{\mu }(x)\right\rangle
	_{-}^{p/\bar{p}},\;\left\langle T_{\mu \nu }(x)\right\rangle ^{p,\bar{p}
	}=\left\langle T_{\mu \nu }(x)\right\rangle _{+}^{p,\bar{p}}+\left\langle
	T_{\mu \nu }(x)\right\rangle _{-}^{p,\bar{p}}\ ,  \notag \\
	\left\langle J^{\mu }(x)\right\rangle _{\pm }^{p/\bar{p}} &=&-ie\,\left. 
	\mathrm{tr}\left[ \gamma ^{\mu }S_{\pm }^{p,\bar{p}}(X,X^{\prime })\right]
	\right\vert _{X=X^{\prime }},\;\left\langle T_{\mu \nu }(x)\right\rangle
	_{\pm }^{p,\bar{p}}=i\,\left. \mathrm{tr}\left[ A_{\mu \nu }S_{\pm }^{p,\bar{
	p}}(X,X^{\prime })\right] \right\vert _{X=X^{\prime }}\ .  
	\label{m8}
\end{eqnarray}

Using representations (\ref{dip01}) and (\ref{dip02}) we obtain that
nonvanishing means are: 
\begin{eqnarray}
	&&\left\langle J^{1}(x)\right\rangle _{+}^{p}=-\left\langle
	J^{1}(x)\right\rangle _{+}^{\bar{p}}=\left\langle J^{0}(x)\right\rangle
	_{+}^{p/\bar{p}},\;\left\langle J^{0}(x)\right\rangle _{+}^{p/\bar{p}}=e\,r^{
	\mathrm{cr}}\left( x-x_{\mathrm{L}}-\frac{K}{\sqrt{eE}}\right) \ ,  \notag \\
	&&\left\langle T^{10}(x)\right\rangle _{+}^{p}=-\left\langle
	T^{10}(x)\right\rangle _{+}^{\bar{p}}=\left\langle T^{00}(x)\right\rangle
	_{+}^{p/\bar{p}},\ \left\langle T^{11}(x)\right\rangle _{+}^{p/\bar{p}
	}=\left\langle T^{00}((x)\right\rangle _{+}^{p/\bar{p}}\ ,  \notag \\
	&&\ \left\langle T^{00}(x)\right\rangle _{+}^{p/\bar{p}}=\frac{r^{\mathrm{cr}
	}}{2}eE\left( x-x_{\mathrm{L}}-\frac{K}{\sqrt{eE}}\right) ^{2},\nonumber\\
	&& \left\langle T^{kk}(x)\right\rangle _{+}^{p/\bar{p}}=\frac{r^{\mathrm{cr}}}{
	2\pi }\log \frac{\sqrt{eE}(x-x_{\mathrm{L}})}{K},\ k=2,\dots ,D,\;\mathrm{
	if\;}\sqrt{eE}(x-x_{\mathrm{L}})\gg K\ ;  \notag \\
	&&\left\langle J^{1}(x)\right\rangle _{-}^{p}=-\left\langle
	J^{1}(x)\right\rangle _{-}^{\bar{p}}=-\left\langle J^{0}(x)\right\rangle ^{p/
	\bar{p}},\;\left\langle J^{0}(x)\right\rangle _{-}^{p/\bar{p}}=-e\,r^{
	\mathrm{cr}}\left( x_{\mathrm{R}}-x-\frac{K}{\sqrt{eE}}\right) \ ,  \notag \\
	&&\left\langle T^{10}(x)\right\rangle _{-}^{p}=-\left\langle
	T^{10}(x)\right\rangle _{-}^{\bar{p}}=-\left\langle T^{00}(x)\right\rangle
	_{-}^{p/\bar{p}},\;\left\langle T^{11}(x)\right\rangle _{-}^{p/\bar{p}
	}=\left\langle T^{00}((x)\right\rangle _{-}^{p/\bar{p}}\ ,  \notag \\
	&&\left\langle T^{00}(x)\right\rangle _{-}^{p/\bar{p}}=\frac{1}{2}r^{\mathrm{
	cr}}eE\left( x_{\mathrm{R}}-x-\frac{K}{\sqrt{eE}}\right) ^{2},\nonumber\\
	&& \left\langle T^{kk}(x)\right\rangle _{-}^{p/\bar{p}}=\frac{r^{\mathrm{cr}}}
	{2\pi }\log  \frac{\sqrt{eE}(x_{\mathrm{R}}-x)}{K},\ k=2,\dots ,D,\;\mathrm{if\;}
	\sqrt{eE}(x_{\mathrm{R}}-x)\gg K\ .  
	\label{emtL1}
\end{eqnarray}
It entails that
\begin{eqnarray}
	\left\langle J^{1}(x)\right\rangle ^{p} &=&-\left\langle
	J^{1}(x)\right\rangle ^{\bar{p}}\approx e\,r^{\mathrm{cr}}L,\ \;\left\langle
	J^{0}(x)\right\rangle ^{p/\bar{p}}\approx 2e\,r^{\mathrm{cr}}x\ ,  \notag \\
	\left\langle T^{10}(x)\right\rangle ^{p} &=&-\left\langle
	T^{10}(x)\right\rangle ^{\bar{p}}=r^{\mathrm{cr}}eELx,\;\left\langle
	T^{11}(x)\right\rangle ^{p/\bar{p}}=\left\langle T^{00}((x)\right\rangle ^{p/
	\bar{p}}\ ,  \notag \\
	\left\langle T^{00}(x)\right\rangle _{-}^{p/\bar{p}} &\approx &r^{\mathrm{cr}
	}eE\left[ \left( \frac{L}{2}\right) ^{2}+x^{2}\right] \ ,  
	\label{emtL2a}
\end{eqnarray}
for $\left\vert x\right\vert <L/2$\ and
\begin{eqnarray}
	\left\langle T^{kk}(x)\right\rangle _{-}^{p/\bar{p}} &\approx &\frac{r^{
	\mathrm{cr}}}{2\pi }\log \left\{ eE\left[ \left( \frac{L}{2}\right)
	^{2}-x^{2}\right] \right\} \;\mathrm{if\;}eE\left[ \left( \frac{L}{2}\right)
	^{2}-x^{2}\right] \gg K^{2}\ ,  \notag \\
	\left\langle T^{kk}(x)\right\rangle _{-}^{p/\bar{p}} &\approx &\frac{r^{
	\mathrm{cr}}}{2\pi }\log \left[ \sqrt{eE}(\frac{L}{2}\pm x)\right] \;\mathrm{%
	if\;}\sqrt{eE}(\frac{L}{2}\mp x)\lesssim K,\ k=2,\dots ,D\ .  
	\label{emtL2b}
\end{eqnarray}

Vacuum polarization contribution $\left\langle J^{\mu }(x)\right\rangle ^{c}$
and $\left\langle T_{\mu \nu }\right\rangle ^{c}$ will be calculated in the
next section. Here we will show how to connect the matrix elements 
(\ref{emtL2a}) and (\ref{emtL2b}) with quantities characterizing 
directly pair production effect.

It should be noted that in strong-field QED with $t$-steps Heisenberg operators 
of physical quantities (for example, the kinetic energy
operator of the Dirac field) are time-dependent in the general case. That is why 
one can determine contributions of the final particles, using \textrm{in-in} 
vacuum means, and setting $t\rightarrow \infty$
(which means considering the time instant when the external field
is already switched off and all the corresponding effects of the vacuum
polarization vanish). In the case under consideration we work with mean
values when they already are time independent and another way of actions has
to be used to determine contributions of the final particles.

We see from Eq. (\ref{emtL1}) that the charge density $\left\langle
J^{0}(x)\right\rangle _{+}^{p/\bar{p}}$, formed by contributions from the
domain $D_{x}^{+}$, is positive while the charge density $\left\langle
J^{0}(x)\right\rangle _{-}^{p/\bar{p}}$, formed by contributions from the
domain $D_{x}^{-}$, is negative. This shows that main contributors to
these densities are the created positrons and electrons, respectively. The
means $\left\langle J^{0}(x)\right\rangle _{+}^{p/\bar{p}}$ grow along the
direction of the electric field as $x\rightarrow x_{\mathrm{R}}$, while the
means $\left\langle J^{0}(x)\right\rangle _{-}^{p/\bar{p}}$ grow in the
opposite direction as $x\rightarrow x_{\mathrm{L}}$. Also the energy density 
$\left\langle T^{00}((x)\right\rangle _{+}^{p/\bar{p}}$ and the pressure
component along the direction of the electric field $\left\langle
T^{11}(x)\right\rangle _{+}^{p/\bar{p}}$, increase as $x\rightarrow x_{
\mathrm{R}}$, while the energy density $\left\langle T^{00}((x)\right\rangle
_{-}^{p/\bar{p}}$ and the pressure $\left\langle T^{11}(x)\right\rangle
_{-}^{p/\bar{p}}$ increase as $x\rightarrow x_{\mathrm{L}}$. Moving along
the direction of the electric field, positrons exit the region 
$S_{\mathrm{int}}$ at $x=x_{\mathrm{R}}$, while electrons moving in the opposite
direction exit the region $S_{\mathrm{int}}$ at $x=x_{\mathrm{L}}$. These
particles maintain directions of their movements after leaving the region 
$S_{\mathrm{int}}$ at $x>x_{\mathrm{R}}$ and $x<x_{\mathrm{L}}$. 
Once outside the region $S_{\mathrm{int}}$, these particles are not affected
by the local effects of vacuum polarization, and cannot change after the
field is turned off. Hence, these are final particles. Their state is
described by $\mathrm{out}$-solutions, given by Eq. (\ref{c17}), see Ref. 
\cite{L-field}. Since the distances $x-x_{\mathrm{L}}$ and 
$x_{\mathrm{R}}-x$ are much larger than the formation length $\Delta l_{0}$ of 
the created pair, one can use the semiclassical description of particle motion.
From this point of view, an electron-positron pair is created with the same
probability at any point inside of the region $S_{\mathrm{int}}$. The
particles are created with a small kinetic energy, which then increases. In
the domains $D_{x}^{+}$ and $D_{x}^{-}$ increments of particle kinetic
energy are, $\left\vert \pi _{0}\left( x\right) \right\vert -\left\vert \pi
_{0}\left( x_{\mathrm{L}}\right) \right\vert =eE\left( x-x_{\mathrm{L}
}\right)$ and $\pi _{0}\left( x\right) -\pi _{0}\left( x_{\mathrm{R}
}\right)=eE\left(x_{\mathrm{R}}-x\right)$ respectively. These particles
are ultrarelativisic, therefore longitudinal momenta of the particles on $x$  
hyperplane are defined by their kinetic energies: $p_{x}\left( x\right)
=\left\vert \pi _{0}\left( x\right) \right\vert$ for positrons with 
$\pi_{0}\left( x\right) \in D_{x}^{+}$ and $p_{x}\left( x\right) =-\pi
_{0}\left( x\right) $ for electrons with $\pi _{0}\left( x\right) \in
D_{x}^{-}$. It is natural to assume that the created particles observed
inside the region $S_{\mathrm{int}}$ near the boundaries $x_{\mathrm{L}}$
and $x_{\mathrm{R}}$ practically do not differ from those observed outside
this region and, therefore, represent the final particles. This situation is
similar to $t$-step case, when final particles are those that remain after
the field is switched off. Thus, we see that fluxes at 
$x\rightarrow x_{\mathrm{L}}$ and at $x\rightarrow x_{\mathrm{R}}$ hyperplanes 
form final particles with energies and momenta from the domains $D_{x}^{+}$ and 
$D_{x}^{-}$.

Thus, it is enough to know longitudinal currents and the energy fluxes
through the surfaces $x=x_{\mathrm{L}}$ and $x=x_{\mathrm{R}}$, given by
Eq. (\ref{emtL2a}), that are formed in the region $S_{\mathrm{int}}$ to
evaluate the contributions of the initial and final states. The normal forms
of the operators $J^{1}$ and $T^{10}$ with respect to the \textrm{out}
-vacuum are:
\begin{equation*}
	N_{\mathrm{out}}\left( J^{1}\right) =J^{1}-\left\langle 0,\mathrm{out}
	\left\vert J^{1}\right\vert 0,\mathrm{out}\right\rangle ,\;\;N_{\mathrm{out}
	}\left( T^{10}\right) =T^{10}-\left\langle 0,\mathrm{out}\left\vert
	T^{10}\right\vert 0,\mathrm{out}\right\rangle \ .
\end{equation*}
Taking into account Eqs. (\ref{4.1b}) and (\ref{emtL2a}), we calculate
densities of the longitudinal current and energy flux corresponding to the
final particles as means with respect to the initial vacuum state,
\begin{eqnarray}
	&&J_{\mathrm{cr}}^{1}(x)=\left\langle N_{\mathrm{out}}\left( J^{1}\right)
	\right\rangle _{\mathrm{in}}=\left\langle J^{1}(x)\right\rangle _{\mathrm{in}
	}-\left\langle J^{1}(x)\right\rangle _{\mathrm{out}}=\nonumber\\
	&&\left\langle
	J^{1}(x)\right\rangle ^{p}-\left\langle J^{1}(x)\right\rangle ^{\bar{p}
	}=2en^{\mathrm{cr}}\,;  \notag \\
	&&T_{\mathrm{cr}}^{10}(x)=\left\langle N_{\mathrm{out}}\left[ T^{10}(x)
	\right] \right\rangle _{\mathrm{in}}=\left\langle T^{10}(x)\right\rangle _{
	\mathrm{in}}-\left\langle T^{10}(x)\right\rangle _{\mathrm{out}
	}=\nonumber\\
	&&\left\langle T^{10}(x)\right\rangle ^{p}-\left\langle
	T^{10}(x)\right\rangle ^{\bar{p}}=2n^{\mathrm{cr}}eEx\ ,  
	\label{m10}
\end{eqnarray}
where $n^{\mathrm{cr}}=r^{\mathrm{cr}}L$ is the total number density of
pairs created per unit time and per unit surface orthogonal to the electric
field direction, whereas $r^{\mathrm{cr}}$ is the pair-production rate,
given by Eq. (\ref{L27}). This rate coincides with the known pair-production
rate in a constant uniform electric field, see Ref. \cite{GavG96a}. Note
that $n^{\mathrm{cr}}$ is proportional to the magnitude of the potential
step $\Delta U\ \mathbb{=\ }eEL$. We stress that the longitudinal current
density $J_{\mathrm{cr}}^{1}(x)$ is $x$-independent. The process of the
current formation has a constant rate per unit length,
\begin{equation}
	\frac{J_{\mathrm{cr}}^{1}(x)}{L}=2e\,r^{\mathrm{cr}}\ .  
	\label{m12}
\end{equation}
The energy flux $T_{\mathrm{cr}}^{10}(x)$ of the final particles through the
surface $x$, is proportional to the potential energy difference with
respect of the hyperplane of the symmetry $x=0$, 
$U\left( x\right) -U\left(0\right) =eEx$ and has the maximal magnitude as 
$x\rightarrow x_{\mathrm{L}} $ and $x\rightarrow x_{\mathrm{R}}$,
\begin{equation}
	T_{\mathrm{cr}}^{10}(x_{\mathrm{R}})=-T_{\mathrm{cr}}^{10}(x_{\mathrm{L}
	})=n^{\mathrm{cr}}\Delta U\ .  
	\label{m16}
\end{equation}
We see that the fluxes of final particles are formed by the fluxes of the
positrons moving along the direction of the electric field and electrons
moving to the opposite direction.

Comparing two nonzero components of the $d$-dimensional Lorentz vectors 
$\left\langle J^{1}(x)\right\rangle ^{p/\bar{p}}$ and $\left\langle
J^{0}(x)\right\rangle ^{p/\bar{p}}$ , given by Eq. (\ref{emtL2a}), we see
the relationship of the charge density of created pairs 
$J_{\mathrm{cr}}^{0}(x)$ with the current densities 
$\left\langle J^{0}(x)\right\rangle ^{p/\bar{p}}$. Namely,
\begin{equation*}
	J_{\mathrm{cr}}^{0}(x)=\left\langle J^{0}(x)\right\rangle ^{p}+\left\langle
	J^{0}(x)\right\rangle ^{\bar{p}}=4e\,r^{\mathrm{cr}}x\ .
\end{equation*}
We see that there exists a charge polarization due to the electric field. In
particular, 
\begin{equation}
	J_{\mathrm{cr}}^{0}(x_{\mathrm{L}})=-2en^{\mathrm{cr}},\;J_{\mathrm{cr}
	}^{0}(x_{\mathrm{R}})=2en^{\mathrm{cr}}\ .  
	\label{m17b}
\end{equation}

A relation of the energy density $T_{\mathrm{cr}}^{00}(x)$ of created pairs
to the mean value $\left\langle T^{00}(X)\right\rangle ^{p/\bar{p}}$ can be
derived in a similar manner as it was done for the current density. For
this, it is suffices to note that the means $\left\langle
T^{00}(X)\right\rangle ^{p/\bar{p}}$ and $\left\langle
T^{10}(X)\right\rangle ^{p/\bar{p}}$ are two nonzero components of a 
$d$-dimensional Lorentz vector. Therefore, by rotating the coordinate system,
we obtain relations between all others diagonal elements of the vacuum mean
values of EMT. These relations are:
\begin{equation}
	T_{\mathrm{cr}}^{\mu \mu }(x)=\left\langle T^{\mu \mu }(x)\right\rangle
	^{p}+\left\langle T^{\mu \mu }(x)\right\rangle ^{\bar{p}}=2\left\langle
	T^{\mu \mu }(x)\right\rangle ^{p}\ ,  
	\label{m18}
\end{equation}
where $\left\langle T^{\mu \mu }(x)\right\rangle ^{p}$ are given by 
Eqs. (\ref{emtL2a}) and (\ref{emtL2b}). In particular, we have
\begin{eqnarray}
	T_{\mathrm{cr}}^{00}(x_{\mathrm{R}}) &=&T_{\mathrm{cr}}^{00}(x_{\mathrm{L}
	})=n^{\mathrm{cr}}\Delta U\ ,  \notag \\
	T_{\mathrm{cr}}^{kk}(x_{\mathrm{R}}) &=&T_{\mathrm{cr}}^{kk}(x_{\mathrm{L}})=
	\frac{r^{\mathrm{cr}}}{\pi }\log \left( \sqrt{eE}L\right) ,\ k=2,\dots ,D\ .
	\label{m19}
\end{eqnarray}

\subsection{Vacuum polarization contributions\label{S52}}

One can verify using Eqs. (\ref{gav11b}) and (\ref{gav13b}) that means 
$\left\langle J^{\mu }(x)\right\rangle ^{c}$ and 
$\left\langle T_{\mu \nu}\right\rangle ^{c}$ are finite as 
$L\rightarrow \infty $. That is why they
can be calculated in such limit as well. In this relation, we recall that
the causal propagator $S^{c}(X,X^{\prime })$ in $L$-constant electric field
was calculated in Ref. (\cite{GF22}) and its limiting expression as 
$L\rightarrow \infty $ was found. Moreover, we have demonstrated that that
expression has the form of the causal propagator $S^{c}(X,X^{\prime })$ in 
$T$-constant electric field as in the limit $T\rightarrow \infty$. In
particular, it was shown that the causal propagator given by 
Eq. (\ref{gav11b}) can be represented in the Schwinger's integral form,
\begin{equation}
	S^{c}(X,X^{\prime })=(\gamma P+m)\Delta ^{c}(X,X^{\prime }),\ \ \Delta
	^{c}(X,X^{\prime })=\int_{0}^{\infty }f(X,X^{\prime };s)ds\ ,  
	\label{gr2a}
\end{equation}
where
\begin{eqnarray}
	&&\ f(X,X^{\prime };s)=\exp \left( -eE\gamma ^{0}\gamma ^{1}s\right)
	\,f^{(0)}(X,X^{\prime };s)\ ,  \notag \\
	&&\ f^{(0)}(X,X^{\prime };s)=-\left( \frac{-i}{4\pi }\right) ^{d/2}\frac{eE\,
	}{s^{(d-2)/2}\sinh (eEs)}  \notag \\
	&&\times \exp \left[ -ism^{2}-\frac{i}{2}eEy_{0}(x+x^{\prime })+\frac{i}{4s}
	\left\vert \mathbf{r}_{\bot }-\mathbf{r}_{\bot }^{\prime }\right\vert ^{2}-
	\frac{i}{4}eE\coth (eEs)\left( y_{0}^{2}-y_{1}^{2}\right) \right]
	\label{gav19b}
\end{eqnarray}
is the Fock-Schwinger kernel \cite{Schwi51}. Here $\ y_{0}=t-t^{\prime}$
and $y_{1}=x^{\prime }-x$. The kernel can be represented as the matrix
element with respect of eigenvectors of the coordinate operators $X^{\mu}$,
\begin{eqnarray}
	&&f(X,X^{\prime };s)=\langle t,\mathbf{r}\mid e^{-isM^{2}}\mid t^{\prime },
	\mathbf{r}^{\prime }\rangle \ ,\;M^{2}=m^{2}-i0-P^{2}-\frac{e}{2}\sigma
	^{\mu \nu }F_{\mu \nu }\ ,  \notag \\
	&&F_{\mu \nu }=\partial _{\mu }A_{\nu }-\partial _{\nu }A_{\mu },\ \ \sigma
	^{\mu \nu }=\frac{i}{2}[\gamma ^{\mu },\gamma ^{\nu }]\,.  
	\label{fc}
\end{eqnarray}
This fact allows one to use results obtained in Ref. \cite{GavGitY12} for
the renormalization of the mean values (\ref{m5.3b}), see details in 
Appendix \ref{Ap3}.

Using this representation, we calculate $\mathrm{Re}\langle j_{\mu }\left(
t\right) \rangle ^{c}$ and $\mathrm{Re}\langle T_{\mu \nu }\left( t\right)
\rangle ^{c}$. It is easy to see that $\langle j_{\mu }\left( t\right)
\rangle ^{c}=0$, as should be expected due to translational symmetry, and 
$\langle T_{\mu \nu }(t)\rangle ^{c}=0\,,\;\mu \neq \nu$.

Let us substitute Eq. (\ref{gr2a}) into Eq. (\ref{m5.3b}) with account taken
of the representation (\ref{gav19b}). Then nonvanishing nonrenormalized
vacuum means can be written as:
\begin{eqnarray*}
	&&\left\langle T_{00}\right\rangle ^{c}=-\left\langle T_{11}\right\rangle
	^{c}=J_{(d)}eE\int_{\Gamma _{c}}\frac{\,f(X,X;s)}{\sinh (eEs)}ds,\
	J_{(d)}=2^{\left\lfloor d/2\right\rfloor -1}\ , \\
	&&\left\langle T_{ii}\right\rangle ^{c}=J_{(d)}eE\int_{\Gamma _{c}}\frac{
	\sinh (eEs)\,}{eEs}f(X,X;s)ds,\ i=2,\dots ,d\ .
\end{eqnarray*}
These means can be expressed via nonrenormalized one-loop Heisenberg-Euler
Lagrangian $\mathcal{L}$, as
\begin{eqnarray}
	&&\left\langle T_{00}\right\rangle ^{c}=-\ \left\langle T_{11}\right\rangle
	^{c}=E\frac{\partial \mathcal{L}}{\partial E}-\mathcal{L},\ \left\langle
	T_{ii}\right\rangle ^{c}=\mathcal{L\ },  \notag \\
	&&\mathcal{L}=\frac{1}{2}\int_{\Gamma _{c}}\frac{ds}{s}\mathrm{tr\,}
	f(X,X;s)\ ,  \notag \\
	&&\mathrm{tr\,}f(X,X;s)=2J_{(d)}\,\cosh (eEs)f^{(0)}(X,X;s)\ .
	\label{unren-emt2}
\end{eqnarray}

To renormalize the mean values (\ref{unren-emt2}) it is enough to
renormalize the real part of the one-loop effective action 
$W=\int \mathcal{L}dtd\mathbf{r}$ (see Ref. \cite{GavGitY12}).

Then we may use the fact that, real parts of the renormalized finite vacuum
mean values are expressed via the renormalized effective Lagrangian 
(\ref{Lren}) as:
\begin{equation}
	\mathrm{Re}\left\langle T_{00}\right\rangle _{\mathrm{ren}}^{c}=-\mathrm{Re}
	\left\langle T_{11}\right\rangle _{\mathrm{ren}}^{c}=E\frac{\partial \mathrm{Re
	}\mathcal{L}_{\mathrm{ren}}}{\partial E}-\mathrm{Re}\mathcal{L}_{\mathrm{ren}
	},\ \mathrm{Re}\left\langle T_{ii}\right\rangle _{\mathrm{ren}}^{c}=\mathrm{Re}
	\mathcal{L}_{\mathrm{ren}}\ .  
	\label{4.4}
\end{equation}

Thus, taking into account Eq. (\ref{n4n}), one can see that in the
strong-field case, the quantities (\ref{4.4}) have the following behavior:
\begin{equation*}
	\mathrm{Re}\left\langle T_{\mu \mu }\right\rangle _{\mathrm{ren}}^{c}\sim
	\left\{ 
	\begin{array}{l}
	\left\vert eE\right\vert ^{d/2},\ d\neq 4n \\ 
	\left\vert eE\right\vert ^{d/2}\log \left( eE/\mu ^{2}\right) ,\ d=4n\ 
	\end{array}
	\right. \ .
\end{equation*}

Finally, we have obtained nonperturbative one-loop representations for the
mean current densities and renormalized EMT of a Dirac field in the 
$L$-constant electric background as:
\begin{eqnarray}
	\left\langle J^{\mu }(x)\right\rangle _{\mathrm{in}} &=&\left\langle J^{\mu
	}(x)\right\rangle ^{p},\;\left\langle J^{\mu }(x)\right\rangle _{\mathrm{out}
	}=\left\langle J^{\mu }(x)\right\rangle ^{\bar{p}}\ ,  \notag \\
	\left\langle T_{\mu \nu }(x)\right\rangle _{\mathrm{in}}^{\mathrm{ren}} &=&
	\mathrm{Re}\left\langle T_{\mu \nu }\right\rangle _{\mathrm{ren}}^{c}+\mathrm{Re}
	\left\langle T_{\mu \nu }(x)\right\rangle ^{p},  \notag \\
	\left\langle T_{\mu \nu }(x)\right\rangle _{\mathrm{out}}^{\mathrm{ren}} &=&
	\mathrm{Re}\left\langle T_{\mu \nu }\right\rangle _{\mathrm{ren}}^{c}+\mathrm{Re}
	\left\langle T_{\mu \nu }(x)\right\rangle ^{\bar{p}}.  
	\label{final-forms}
\end{eqnarray}
Here $\mathrm{Re}\left\langle T_{\mu \nu }\right\rangle _{\mathrm{ren}}^{c}$ 
is given by Eq. (\ref{4.4}), other terms are related by Eq. (\ref{emtL2a})
and (\ref{emtL2b}), and expressed via characteristics of pair creation as
$\langle J^{\mu}(x)\rangle ^{p} = J_{\textrm{cr}}^\mu(x)/2$ and   
$\textrm{Re}\langle T^{\mu\nu}(x)\rangle ^{p}= T_{\textrm{cr}}^{\mu\nu}(x)/2$. 
The components $\mathrm{Re}\left\langle T_{\mu \nu }\right\rangle_{\mathrm{ren}
}^{c}$ describe the contribution due to vacuum polarization. These
components are local. The components $J_{\mathrm{cr}}^{\mu }(x)$ and 
$T_{\mathrm{cr}}^{\mu \nu }(x)$ describe the contribution due to the creation of
real particles from vacuum. They are global quantities and growing unlimited
as the magnitude of potential energy tends to infinity.

\section{Discussion and summary\label{SConcl}}

In this work we draw the reader's attention to the fact that the technique
of nonperturbative calculating of vacuum instability effects based on the
original formulation of the strong-field QED with $x$-electric steps
proposed in Refs. \cite{GavGi16,GavGi20} must be refined and supplemented by
a certain regularization procedure studying the problem of local mean
values, see Sec. \ref{S2a}. Here we illustrate general considerations by the
case of strong-field QED with $L$-constant field (which can be interpreted
as an electric field between capacitor plates). In the same case, we propose
a convenient volume regularization procedure with respect of the
time-independent inner product on the $t$-constant hyperplane. At the same
time we find adequate representations (\ref{gav11b}), (\ref{gav11c}), and 
(\ref{gav13a}) for all the involved singular spinor functions. Using the
regularization procedure and the singular functions, we calculate the vacuum
mean values of current density and EMT (\ref{m5.3b}) that are local physical
quantities. The new approach allows us to separate in these mean values
global contributions due to the particle creation from local contributions
due to the vacuum polarization. In Sec. \ref{S52} we show that real parts of
the vacuum polarization contributions to EMT can be expressed via the
renormalized effective Heisenberg--Euler Lagrangian. Finally, we have
obtained nonperturbative one-loop representations for the mean current
densities and renormalized EMT (\ref{final-forms}).

It's believed that in the limiting case $L\rightarrow \infty$ the 
$L$-constant field is a suitable regularization of the constant uniform
electric field in course of describing the vacuum instability effects when
the field region is considered to be small compared to the entire field
region and far enough from its boundaries. In this relation, it is
demonstrated that the longitudinal current density of created particles 
$J_{\mathrm{cr}}^{1}(x)=2en^{\mathrm{cr}}$ is $x$-independent and the process of
the current formation has a constant rate per unit length (\ref{m12}) that
coincides with the known pair-production rate in a constant uniform electric
field. This fact confirms the above supposition and justifies the proposed
regularization procedure (\ref{renorm}).

The new approach applied to study the vacuum instability in the $L$-constant
field allows us to reveal details that could not be detected by calculations
in the homogeneous electric field. For example, the obtained formulas show
explicitly that the current density and EMT of created particles are formed
separately by contributions of created positrons and created electrons. The
behavior of these quantities can be described as follows. They grow with the
increase the potential energy differences with respect of the symmetry
hyperplane $x=0$ and reach maximal magnitudes near the capacitor plates,
namely as $x\rightarrow x_{\mathrm{L}}$ and as 
$x\rightarrow x_{\mathrm{R}}$. They are growing unlimited as the magnitude of the 
potential energy $\Delta U$ tends to infinity. Note that it explains initiation 
of secularly
growing loop corrections to two-point correlation functions in the case of
the time-independent electric field given by a linear potential step 
\cite{AkhmP15}. The longitudinal energy flux of final particles on both
sides of the hyperplane $x=0$ is directed in opposite directions and a
charge polarization occurs due to the electric field. Continuing to move
along the direction of the electric field, the positrons leave the field at
the point $x=x_{\mathrm{R}}$, and the electrons moving in the opposite
direction leave the field at the point $x=x_{\mathrm{L}}$. The current
density and EMT calculated for these separated fluxes of electrons and
positrons \cite{GavGi20} add up to results that are consistent with the
results obtained in this article.

It is useful to compare the obtained results with results on the study of
the vacuum instability in the $L$-constant electric field presented in the
work \cite{GavGi20}. In the latter work, it was calculated the current
densities and the energy flux densities of electrons and positrons, after the
instant when these fluxes become completely separated and have
left the region $S_{\mathrm{int}}$ . In the framework of the approach
formulated in the present article, it is impossible to consider processes of
particles leaving the region $S_{\mathrm{int}}$ boundaries. Nevertheless,
based on physical considerations, we can expect a certain agreement between
both results. In particular, we see that current density and EMT components
given by Eqs. (\ref{m10}), (\ref{m16}), (\ref{m17b}), and (\ref{m19}) for 
$x\rightarrow x_{\mathrm{L}}$ and $x\rightarrow x_{\mathrm{R}}$ are sums of
the corresponding values obtained in Ref. \cite{GavGi20} separately for
electrons and positrons. This may be considered as an additional evidence
that the proposed renormalization procedure (\ref{renorm}) is consistent.

We believe also that results obtained in this work may contribute to a
further development of the locally constant field approximation which is not
based on the Heisenberg--Euler Lagrangian approach.

\section{Acknowledgement}

The work is supported by Russian Science Foundation, grant No. 19-12-00042.

\section{Appendix}

\subsection{Decomposition of observavel into plane waves \label{Ap1}}

In the framework of a field theory an observable $F$ can be realized as an
inner product of the type (\ref{t4}) of localizable wave packets $\psi (X)$
and $\hat{F}\psi ^{\prime }(X)$,
\begin{equation*}
	F\left( \psi ,\psi ^{\prime }\right) =\left( \psi ,\hat{F}\psi ^{\prime
	}\right) \ ,
\end{equation*}
where $\hat{F}$ is a differential operator and $\psi (X)$ and 
$\psi^{\prime }(X)$ are solutions of the Dirac equation. Assuming that an
observable $F\left( \psi ,\psi ^{\prime }\right) $ is time-independent
during the time $T$\ one can represent this observable in the following form
of an average value over the period $T$:
\begin{equation*}
	\left\langle F\right\rangle =\frac{1}{T}\int_{-T/2}^{+T/2}F\left( \psi ,\psi
	^{\prime }\right) dt\ .
\end{equation*}
In general the wave packets $\psi (X)$ and $\psi ^{\prime }(X)$ can be
decomposed into plane waves $\psi _{n}(X)$ and $\psi _{n}^{\prime }(X)$
with given $n$, 
\begin{equation*}
	\psi (X)=\sum_{n}\alpha _{n}\psi _{n}(X),\;\psi ^{\prime }(X)=\sum_{n}\alpha
	_{n}^{\prime }\psi _{n}^{\prime }(X)\ ,
\end{equation*}
where $\psi _{n}(X)$ and $\psi _{n}^{\prime }(X)$ are superpositions of
the solutions$_{\;\zeta }\psi _{n}\left( X\right)$ and 
$^{\;\zeta }\psi_{n}\left( X\right)$. 
Taking into account the orthogonality relation (\ref{c3b}) one finds that the 
decomposition of $\left\langle F\right\rangle$ into plane waves with given 
$n$ does not contain interference terms,
\begin{equation*}
	\left\langle F\right\rangle =\sum_{n}F\left( \alpha _{n}\psi _{n},\alpha
	_{n}^{\prime }\psi _{n}^{\prime }\right) \ .
\end{equation*}

\subsection{Integrals on $t$-constant hyperplane\label{Ap2}}

Integrating in (\ref{t4}) over the coordinates $\mathbf{r}_{\bot }$ and
using the structure of constant spinors $v_{\sigma }$ that enter the states 
$\psi_{n}\left( X\right)$ and $\psi_{n^{\prime }}^{\prime }\left( X\right)$, 
we obtain:
\begin{eqnarray*}
	&&\left( \psi _{n},\psi _{n^{\prime }}^{\prime }\right) =\delta _{\sigma
	,\sigma ^{\prime }}\delta _{\mathbf{p}_{\bot },\mathbf{p}_{\bot }^{\prime
	}}V_{\bot }\mathcal{R},\ \ \mathcal{R}=\int_{-K^{\left( \mathrm{L}\right)
	}}^{K^{\left( \mathrm{R}\right) }}\Theta dx\ , \\
	&&\Theta =e^{i\left( p_{0}-p_{0}^{\prime }\right) t}\varphi _{n}^{\ast
	}\left( x\right) \left[ p_{0}+p_{0}^{\prime }-2U\left( x\right) \right]
	\left[ p_{0}^{\prime }-U\left( x\right) +i\partial _{x}\right] \varphi
	_{n^{\prime }}^{\prime }\left( x\right) \ .
\end{eqnarray*}
Then we represent the integral $\mathcal{R}$ as follows
\begin{equation}
	\mathcal{R}=\int_{-K^{\left( \mathrm{L}\right) }}^{-k^{\left( \mathrm{L}
	\right) }}\Theta dx+\int_{-k^{\left( \mathrm{L}\right) }}^{k^{\left( \mathrm{
	R}\right) }}\Theta dx+\int_{k^{\left( \mathrm{R}\right) }}^{K^{\left( 
	\mathrm{R}\right) }}\Theta dx\ ,  
	\label{i2}
\end{equation}
where $0<k^{\left( \mathrm{L}\right) }\ll K^{\left( \mathrm{L}\right) }$
 and $0<k^{\left( \mathrm{R}\right) }\ll K^{\left( \mathrm{R}\right)}$. 
 Parameters $k^{\left( \mathrm{L,R}\right) }$ are selected so
that one can use the asymptotic behavior of WPCF's with large $\left\vert
\xi \right\vert $. It can be seen that for a particular case of the plane
waves with equal $n$ and $\varphi _{n}^{\prime }\left(
x\right) =\varphi _{n}\left( x\right) $ the kernel $\Theta $ is real
constant. Therefore that integrals over intervals $\left[ -K^{\left( \mathrm{
L}\right) },-k^{\left( \mathrm{L}\right) }\right] $ and $\left[ k^{\left( 
\mathrm{R}\right) },K^{\left( \mathrm{R}\right) }\right] $ in Eq. (\ref{i2})
are proportionate to lengths of these intervals. In the case of a
sufficiently large length $L$, one can assume that both lengths $K^{\left( 
\mathrm{L}\right) }-k^{\left( \mathrm{L}\right) }$ and $K^{\left( \mathrm{R}
\right) }-k^{\left( \mathrm{R}\right) }$ are large too, it is of order of
length $L$ and much larger than interval $k^{\left( \mathrm{R}\right)
}+k^{\left( \mathrm{L}\right) }$. In this case the contribution to the
integral (\ref{i2}) from last interval can be ignored. Thus, the value of
the integral (\ref{i2}) is basically determined by the first and last terms.
To calculate these integrals, it is enough to use the asymptotic behavior of
WPCF's both in region with $\xi <0$ and in the region with $\xi >0$. Note
that in intervals $\left[ -K^{\left( \mathrm{L}\right) },-k^{\left( \mathrm{L
}\right) }\right] $ and $\left[ k^{\left( \mathrm{R}\right) },K^{\left( 
\mathrm{R}\right) }\right] $ the modulus of a longitudinal momentum is well
defined as $\left\vert p_{x}\left( x\right) \right\vert =\sqrt{
\left[ \pi _{0}\left( x\right) \right] ^{2}-\pi _{\bot }^{2}}$. Further 
calculation is no different from what is given in the Appendix B in
Ref. \cite{GavGi16}. In particular, one sees that all the solutions 
$_{\;\zeta }\psi _{n}\left( X\right) $ and $^{\;\zeta }\psi _{n}\left(
X\right)$ having different quantum numbers $n$ are orthogonal with respect
to the introduced inner product on the hyperplane $t=\mathrm{const}$. One
can see that the norms of the solutions $_{\;\zeta }\psi _{n}\left( X\right) 
$ and $^{\;\zeta }\psi _{n}\left( X\right)$ with respect to the inner product 
(\ref{t4}) are proportional to the macroscopically large parameters
\begin{equation*}
	\tau ^{\left( \mathrm{L}\right) }=K^{\left( \mathrm{L}\right) }/v^{\mathrm{L}
	},\;\tau ^{\left( \mathrm{R}\right) }=K^{\left( \mathrm{R}\right) }/v^{
	\mathrm{R}},
\end{equation*}
where $v^{\mathrm{L}}=\left\vert p_{x}\left( x\right) /\pi
_{0}\left( x\right) \right\vert \rightarrow c$ and $v^{\mathrm{R}}=v^{
\mathrm{L}}=\left\vert p_{x}\left( x\right) /\pi _{0}\left( x\right)
\right\vert \rightarrow c$ ($c=1$) are absolute values of longitudinal
velocities of particles in the spatial regions where $\left\vert \xi
\right\vert $ is large.

Finally, one obtains the orthonormality relations (\ref{i12}).

\subsection{Ultraviolet renormalization\label{Ap3}}

The one-loop effective action $W=\int \mathcal{L}dtd\mathbf{r}$ can be
represented as $W=(-i/2)\ln \det M^{2}$. After passing to the Euclidean
metric
\begin{equation*}
	t\rightarrow -i\eta ,\quad \partial _{t}\rightarrow i\partial _{\eta },\quad
	eE\rightarrow -iB,\quad B>0\ ,
\end{equation*}
$M^{2}$ becomes the elliptic operator $\tilde{M}^{2}$ and $W$ becomes the
effective action $\tilde{W}=-i\left[ \int \mathcal{L}d\eta d\mathbf{r}\right]
_{qE\rightarrow iB}$\ over the Euclidean space. To carry out the
renormalization procedure, we first introduce the generalized zeta function
of the operator $\tilde{M}^{2}$ in $d$-dimensional Euclidean space using the
heat kernel $K(u)$:
\begin{eqnarray}
	&&\zeta ^{(d)}(s)=\frac{1}{\Gamma (s)}\int_{0}^{\infty }u^{s-1}du\,K(u)\ , 
	\notag \\
	&&K(u)=\int d\eta d\mathbf{r\,}\mathrm{tr\,}f_{\emph{Eucl}}(X,X;u)\ ,  \notag
	\\
	&&f_{\emph{Eucl}}(X,X;u)=\left\langle \eta ,\mathbf{r}\right. \left\vert
	\exp \left( -u\mu ^{-2}\tilde{M}^{2}\right) \right\vert \left. \eta ,\mathbf{
	r}\right\rangle \ .  
	\label{fec}
\end{eqnarray}
Here $\mu$ is a normalization constant with the mass dimension which is necessary 
for the generalized zeta function to be dimensionless. One can write:
\begin{equation*}
	\ln \det \tilde{M}^{2}=\mathrm{tr}\ln \tilde{M}^{2}=-\left. \frac{d\zeta
	^{(d)}(s)}{ds}\right\vert _{s=0}\ .
\end{equation*}

The renormalized effective Lagrangian can be expressed in terms of the
generalized zeta function as:
\begin{equation}
	\mathrm{Re}\mathcal{L}_{\mathrm{ren}}=\mathrm{Re}\left. \mathcal{\tilde{L}}
	\right\vert _{B=ieE},\quad \mathcal{\tilde{L}=-}\frac{1}{2\Omega _{(d)}}
	\left. \frac{d\zeta ^{(d)}(s)}{ds}\right\vert _{s=0},\quad \Omega
	_{(d)}=\int d\eta d\mathbf{r}\ .  
	\label{Lren}
\end{equation}
Next, we calculate $\mathcal{L}_{\mathrm{ren}}$. Comparing Eq. (\ref{fec})
with Eq. (\ref{fc}) and given the expression (\ref{gav19b}) for the kernel,
we calculate the trace,
\begin{eqnarray*}
	&&\mathrm{tr\,}f_{\emph{Eucl}}(X,X;u)=-\left. \mathrm{tr\,}f\left( X,X;-
	\frac{iu}{\mu ^{2}}\right) \right\vert _{eE\rightarrow -iB} \\
	&=&2J_{(d)}\frac{Bu}{\mu ^{2}}\left( \frac{\mu ^{2}}{4\pi u}\right)
	^{d/2}\coth \left( \frac{Bu}{\mu ^{2}}\right) \exp \left[ -\left( \frac{m}{
	\mu }\right) ^{2}u\right] \ .
\end{eqnarray*}
Then for the zeta function in two dimensions we obtain:
\begin{equation*}
	\zeta ^{(2)}(s)=\frac{\Omega _{(2)}}{2\pi \Gamma (s)}\int_{0}^{\infty
	}u^{s-1}B\coth \left( \frac{Bu}{\mu ^{2}}\right) \exp \left[ -\left( \frac{m
	}{\mu }\right) ^{2}u\right] du\ .
\end{equation*}
The integral over $u$ can be expressed in terms of the Hurwitz zeta function
as follows:
\begin{equation*}
	\zeta _{\mathrm{H}}(s,a)=\sum_{k=0}^{\infty }(k+a)^{-s},\quad \mathrm{Re}s>1\ ,
\end{equation*}
whose analytic continuation to the entire complex plane can be given by the
integral representation
\begin{equation*}
	\zeta _{\mathrm{H}}(s,a)=\frac{1}{\Gamma (s)}\int_{0}^{\infty }x^{s-1}\frac{
	e^{-ax}}{1-e^{-x}}dx\ .
\end{equation*}
Then for zeta function $\zeta ^{(2)}(s)$ we obtain (see Refs. \cite{blau1991}, 
\cite{El1994}): 
\begin{equation}
	\zeta ^{(2)}(s)=\left\{ 
	\begin{array}{l}
	\frac{\Omega _{(2)}B}{2\pi }\left[ 2\left( \frac{2B}{\mu ^{2}}\right)
	^{-s}\zeta _{\mathrm{H}}\left( s,1+\frac{m^{2}}{2B}\right) +\left( \frac{
	m^{2}}{\mu ^{2}}\right) ^{-s}\right] ,\quad m\neq 0 \\ 
	\Omega _{(2)}\frac{B}{\pi }\left( \frac{2B}{\mu ^{2}}\right) ^{-s}\zeta _{
	\mathrm{R}}\left( s\right) ,\quad m=0
	\end{array}
	\right. \ .  
	\label{z2repr}
\end{equation}
For $d>2$, the zeta function $\zeta ^{(d)}(s)$ can be expressed in terms of
the function $\zeta ^{(2)}(s)$ as follows:
\begin{equation}
	\zeta ^{(d)}(s)=J_{(d)}\frac{\Omega _{(d)}}{\Omega _{(2)}}\left( \frac{\mu
	^{2}}{4\pi }\right) ^{d/2-1}\frac{\Gamma (s-d/2+1)}{\Gamma (s)}\zeta
	^{(2)}(s-d/2+1),\quad d>2\ .  
	\label{zdrepr2}
\end{equation}
Using Eq. (\ref{Lren}), we obtain the real part of $\mathcal{L}_{\textrm{ren}}$ 
for an arbitrary dimension $d$ in the following form:
\begin{eqnarray*}
	&&\mathrm{Re}\mathcal{L}_{\mathrm{ren}}=\mathcal{-}\frac{1}{2\Omega _{(d)}}
	\mathrm{Re}\left. \frac{d\zeta ^{(d)}(s)}{ds}\right\vert _{s=0,\,B=ieE} \\
	&=&\mathcal{-}\frac{J_{(d)}}{2\Omega _{(2)}}\left( \frac{\mu ^{2}}{4\pi}
	\right) ^{d/2-1}\mathrm{Re}\left. \frac{d}{ds}\left\{ \frac{\Gamma (s-d/2+1)}{
	\Gamma (s)}\zeta ^{(2)}(s-d/2+1)\right\} \right\vert _{s=0,\,B=ieE}\ .
\end{eqnarray*}

The derivative of the zeta function $\zeta ^{(d)}(s)$ at the point $s=0$
reads:
\begin{equation}
	\left. \frac{d\zeta ^{(d)}(s)}{ds}\right\vert _{s=0}=J_{(d)}\left( \frac{\mu
	^{2}}{4\pi }\right) ^{d/2-1}\left\{ 
	\begin{array}{l}
	\frac{(-1)^{d/2-1}}{\Gamma (d/2)}\left[ \zeta ^{(2)^{\prime }}\left( 1-\frac{
	d}{2}\right) +(\gamma +\psi (\frac{d}{2}))\zeta ^{(2)}\left( 1-\frac{d}{2}%
	\right) \right] ,\ \emph{for\,even\,d} \\ 
	\Gamma \left( 1-\frac{d}{2}\right) \zeta ^{(2)}\left( 1-\frac{d}{2}\right)
	,\ \emph{for\,odd\ d}
	\end{array}
	\right. \ .  
	\label{dzd}
\end{equation}
For odd $d$, Eq. (\ref{dzd}) implies that
\begin{equation}
	\left. \frac{d\zeta ^{(d)}(s)}{ds}\right\vert _{s=0}=\left. \Gamma (s)\zeta
	^{(d)}(s)\right\vert _{s=0},\ \ \emph{for\,odd\ d}\ .  
	\label{f58}
\end{equation}

The corresponding final expressions for $\mathcal{\tilde{L}}$ in $d=2$,$3$,
$4$ dimensions are treated in detail in Ref. \cite{blau1991}. For example,
for $d=4$ and $m\neq 0$ we obtain:
\begin{eqnarray*}
	&&\mathcal{\tilde{L}}_{d=4}=\Omega _{(4)}\left( \frac{B}{\pi }\right)
	^{2}\times\\
	&&\left\{ \left[ \log \frac{2B}{\mu ^{2}}-1\right] \zeta _{H}\left( -1,1+
	\frac{m^{2}}{2B}\right) -\left. \frac{\partial }{\partial s}\zeta _{H}\left(
	s-1,1+\frac{m^{2}}{2B}\right) \right\vert _{s=0}+\frac{m^{2}}{2B}\left( \log 
	\frac{m}{\mu }-\frac{1}{2}\right) \right\} \ , \\
	&&\mathrm{Re}\mathcal{L}_{\mathrm{ren}}^{d=4}=\left. \mathrm{Re}\mathcal{\tilde{L
	}}_{d=4}\right\vert _{B=ieE}\ .
\end{eqnarray*}
In particular, in $d=3$ dimension and for $m\neq 0$ we have:
\begin{equation*}
	\mathrm{Re}\mathcal{L}_{\mathrm{ren}}^{d=3}=\mathrm{Re}\left\{ \frac{B}{2\pi }
	\sqrt{2B}\left[ \zeta _{H}\left( -\frac{1}{2},\frac{m^{2}}{2B}+1\right) +
	\frac{1}{2}\sqrt{\frac{m^{2}}{2B}}\right] _{B=ieE}\right\} \ .
\end{equation*}
Using the relation (see Ref. \cite{blau1991}, formula (4.5))
\begin{equation*}
	\zeta _{H}\left( -\frac{1}{2},\frac{m^{2}}{2B}+1\right) =\zeta _{R}\left( -
	\frac{1}{2}\right) -\sum_{l=1}^{\infty }(-1)^{l}\frac{(2l-3)!!}{2^{l}l!}
	\left( \frac{m^{2}}{2B}\right) ^{l}\zeta _{R}\left( l-\frac{1}{2}\right) \ ,
\end{equation*}
we obtain for small $m^{2}/(2\left\vert eE\right\vert)$ the following
result:
\begin{eqnarray*}
	&&\mathrm{Re}\mathcal{L}_{\mathrm{ren}}^{d=3}=\frac{1}{8\pi ^{2}}\left(
	eE\right) ^{3/2}\zeta _{R}\left( \frac{3}{2}\right) +\frac{m^{2}}{8\pi }
	\sqrt{eE}\zeta _{R}\left( \frac{1}{2}\right) \\
	&&-\frac{m^{3}}{4\pi }\sum_{l=2}^{\infty }(-1)^{l}\frac{\sin \frac{\pi l}{2}
	-\cos \frac{\pi l}{2}}{\sqrt{2}}\frac{(2l-3)!!}{2^{l}l!}\left( \frac{m^{2}}{
	2eE}\right) ^{l-3/2}\zeta _{R}\left( l-\frac{1}{2}\right) \ .
\end{eqnarray*}
Note that when the field is very strong, $m^{2}/(eE)\ll 1$, the main
contributions to $\mathcal{L}_{\mathrm{ren}}$ are determined by Eq. 
(\ref{Lren}) as $m\rightarrow 0$. Using (\ref{z2repr}), (\ref{zdrepr2}) and 
(\ref{f58}) for these contributions, we obtain:
\begin{equation*}
	\mathrm{Re}\mathcal{L}_{\mathrm{ren}}\approx \frac{1}{2}\mathrm{Re}\left\{ 
	\begin{array}{l}
	\left[ \log \left( \frac{B}{\mu ^{2}}\right) \frac{\zeta ^{(d)}(0)}{\Omega
	_{(d)}}\right] _{_{B=ieE}},\quad \ \mathrm{for\,even}\emph{\,}d \\ 
	-\left[ \left. \Gamma (s)\frac{\zeta ^{(d)}(s)}{\Omega _{(d)}}\right\vert
	_{s=0}\right] _{_{B=ieE}},\ \ \mathrm{for\,odd}\emph{\,}d
	\end{array}
	\right. \ .
\end{equation*}
In particular,
\begin{equation*}
	\mathrm{Re}\mathcal{L}_{\mathrm{ren}}\approx -\left\{ 
	\begin{array}{l}
	\mathrm{Re}\left[ \frac{B}{4\pi }\log \left( \frac{B}{\mu ^{2}}\right) \right]
	_{B=ieE}=-\frac{eE}{8},\quad \ \,d=2 \\ 
	\mathrm{Re}\left[ \frac{1}{2\pi ^{2}}\left( \frac{B}{2}\right) ^{3/2}\zeta
	_{R}\left( \frac{3}{2}\right) \right] _{B=ieE}=-\frac{\left( eE\right) ^{3/2}
	}{8\pi ^{2}}\zeta _{R}\left( \frac{3}{2}\right) ,\ d=3 \\ 
	\mathrm{Re}\left[ \frac{1}{2}\Omega _{(4)}\zeta ^{\prime (4)}(0)\right]
	_{B=ieE}\approx \frac{(eE)^{2}}{24\pi ^{2}}\log \left( \frac{eE}{\mu ^{2}}
	\right) ,\ \ d=4
	\end{array}
	\right. \ .
\end{equation*}
In the general case, for a very strong electric field, we have:
\begin{equation}
	\mathrm{Re}\mathcal{L}_{\mathrm{ren}}\sim \left\{ 
	\begin{array}{l}
	\left\vert eE\right\vert ^{d/2},\quad d\neq 4n\  \\ 
	\left\vert eE\right\vert ^{d/2}\log \left( eE/\mu ^{2}\right) ,\quad d=4n
	\end{array}
	\right. \ .  
	\label{n4n}
\end{equation}


\end{document}